\documentstyle[prl,aps,floats,epsfig]{revtex}
\def\be{\begin{eqnarray}}
\def\ee{\end{eqnarray}}
\def\lsim{\lesssim}

\def\etal{{\it et al.}}
\def\del{\partial}
\def\vA{{\bbox A}}
\def\vI{{\bbox I}}

\def\vV{{\bbox V}}
\def\vr{{\bbox r}}
\def\vq{{\bbox q}}
\def\vk{{\bbox k}}
\def\vp{{\bbox p}}
\def\vbp{{\overline {\bbox p}}}
\def\vx{{\bbox x}}
\def\vs{{\bbox \sigma}}
\def\vzero{{\bbox 0}}
\def\hatr{{\bbox {\hat r}}}
\def\vtau{{\vec \tau}}
\newcommand{\e}{{\mbox{e}}}
\def\fm{{\mbox{fm}}}
\def\MeV{{\mbox{MeV}}}
\def\ve{\varepsilon}

\def\He#1{{}^{#1}\mbox{He}}

\def\nlo#1{\mbox{N$^{#1}$LO}}
\def\Sunit{\mbox{$10^{-20}$ keV-b}}
\def\calM{{\cal M}}
\def\calR{{\cal R}}
\def\oO{{\cal O}}
\def\oT{{\cal T}}
\def\voO{{\bbox {\cal O}}}
\def\voT{{\bbox {\cal T}}}
\def\okin{{\cal O}_{{\cal P}}}
\def\vokin{\bbox{{\cal O}}_{{\cal P}}}
\def\dR{{\hat d^R}}
\def\bi{\bibitem}
\def\prl{Phys. Rev. Lett.}
\def\np{Nucl. Phys.}
\def\pr{Phys. Rev.}
\def\pl{Phys. Lett.}
\def\zint{{\int_0^1\!dz\,}}
\def\zz{{z {\bar z}}}
\def\bLambda{{\bar \Lambda}}
\begin{document}

\twocolumn[%
\hsize\textwidth\columnwidth\hsize\csname@twocolumnfalse\endcsname

\renewcommand{\thefootnote}{\fnsymbol{footnote}}
\setcounter{footnote}{0}

 \hskip 15cm KIAS P02049

 \hskip 15cm Saclay T02/102
 \vskip 1cm
\title{%
Parameter-free effective field theory calculation \\ for the solar
proton-fusion and  $hep$ processes}

\author{%
{\bf T.-S. Park}$^{(a, b)}$, {\bf  L.E. Marcucci}$^{(c,d)}$, {\bf
R. Schiavilla}$^{(e,f)}$, {\bf M. Viviani}$^{(d,c)}$, {\bf A.
Kievsky}$^{(d,c)}$, {\bf S. Rosati}$^{(d,c)}$
\\
{\bf K. Kubodera}$^{(a,b)}$, {\bf D.-P. Min}$^{(g)}$, and {\bf M.
Rho}$^{(a,h)}$ }
\address{%
(a) School of Physics, Korea Institute for Advanced Study, Seoul
130-012, Korea\\
(b) Department of Physics and Astronomy,
University of South Carolina, Columbia, SC 29208, USA\\
(c) Department of Physics, University of Pisa,
I-56100 Pisa, Italy\\
(d) INFN, Sezione di Pisa, I-56100 Pisa, Italy\\
(e) Department of Physics, Old Dominion University,
         Norfolk, Virginia 23529, USA \\
(f) Jefferson Lab, Newport News, Virginia 23606, USA \\
(g) Department of Physics, Seoul National University, Seoul
151-742, Korea\\
(h) Service de Physique Th\'{e}orique, CEA/DSM/SPhT, Unit\'e de
recherche associ\'e au CNRS, CEA/Saclay, 91191 Gif-sur-Yvette
C\'edex, France}

\maketitle

\centerline{(\today)}

\begin{abstract}
 Spurred by the recent complete determination of the weak
currents in two-nucleon systems up to ${\cal O}(Q^3)$ in
heavy-baryon chiral perturbation theory, we carry out a
parameter-free calculation of the threshold $S$-factors for the
solar $pp$ (proton-fusion) and $hep$ processes in an effective
field theory that {\it combines} the merits of the standard
nuclear physics method and systematic chiral expansion. The power
of the EFT adopted here is that one can correlate in a unified
formalism the weak-current matrix elements of two-, three- and
four-nucleon systems. Using the tritium $\beta$-decay rate as an
input to fix the only unknown parameter in the theory, we can
evaluate the threshold $S$ factors with drastically improved
precision; the results are $S_{pp}(0) = 3.94\!\times\!(1 \pm
0.004) \!\times\!10^{-25}\ \mbox{MeV-b}$ and $S_{hep}(0) = (8.6
\pm 1.3)\!\times\! 10^{-20} \ \mbox{keV-b}$. The dependence of
the calculated $S$-factors on the momentum cutoff parameter
$\Lambda$ has been examined for a physically reasonable range of
$\Lambda$. This dependence is found to be extremely small for the
$pp$ process, and to be within acceptable levels for the $hep$
process, substantiating the consistency of our calculational
scheme.

\end{abstract}

 \vskip1pc]

\renewcommand{\thefootnote}{\arabic{footnote}}
\setcounter{footnote}{0}

\section{Introduction}
 The standard approach to nuclear physics~\cite{snpa}
anchored on wavefunctions obtained from the Schr\"odinger (or
Lippman-Schwinger) equation with ``realistic" phenomenological
potentials has scored an impressive quantitative success in
describing systems with two or more nucleons, achieving in some
cases accuracy that defies the existing experimental precision. We
refer to this approach as SNPA (standard nuclear physics
approach). The advent of quantum chromodynamics (QCD) as $the$
theory of strong interactions raises a logical question: What is
the status of SNPA in the context of the fundamental theory, QCD ?
Put more bluntly, is SNPA (despite its great success) just a
model-dependent approach unrelated to the fundamental theory ? In
our view this is one of the most important issues in nuclear
physics today. In this paper, we propose that SNPA can be properly
identified as a $legitimate$ component in the general edifice of
QCD. The next important question is: If SNPA is indeed a bona-fide
element of QCD, how can we establish an expansion scheme which
includes SNPA as an approximation and which allows a systematic
calculation of correction terms with error estimation ? This
systematic correction with an error estimation is not feasible
with SNPA alone.

Broadly speaking, formulating nuclear physics calculations starting
from effective field theories (EFTs) based on QCD
calls for ``double-step decimation"~\cite{doubledecimation}.
To start with,
the ``bare" EFT Lagrangian for strong interaction
physics in the nonperturbative sector is defined at the chiral scale
$\Lambda_\chi\sim 4\pi f_\pi\sim 1$ GeV, with the parameters in
the Lagrangian determined by suitably matching them
to QCD at that
scale~\cite{harada-yamawaki}. This Lagrangian cannot be directly
used for studying complex nuclei. In order to
render it applicable to low-energy nuclear physics, it
is desirable, if not indispensable, that a ``decimation" be made
from the chiral scale $\Lambda_\chi$ down to what might be called
the ``Fermi-sea scale" $\Lambda_{FS}\sim k_F$,
where $k_F$ the nuclear Fermi momentum.
An effective Lagrangian resulting from the
decimation to this Fermi-sea scale is expected to contain
parameters that reflect ``intrinsic" density-dependence
(as suggested in, {\it e.g.},Ref.~\cite{BR})
{\it in addition} to the usual dense loop
effects. The next stage of decimation consists of integrating out
the effective degrees of freedom and modes from the scale
$\Lambda_{FS}$ to $\Lambda_0\sim 0$ MeV. This stage corresponds to
doing shell-model type calculations in finite
nuclei~\cite{kuo} and to formulating Fermi-liquid theory in
nuclear matter~\cite{schwenk}.
The physics of heavy nuclei or
nuclear matter will involve both decimations but, for light nuclei,
one can bypass the second decimation and work directly with the
chiral Lagrangian.

 The aim of this article is to
describe a formalism
that combines the high accuracy of
SNPA and the power of EFT to make totally parameter-free {\it
predictions} for electroweak transitions in light nuclei.
To be concrete, we shall consider the following
two solar nuclear fusion processes:
\be pp:&&\ \ \
p+p\rightarrow d + e^+ +\nu_e\,, \label{pp}
\\
hep:&&\ \ \ p+\He3 \rightarrow \He4 + e^+ + \nu_e\,. \label{hep}
\ee
We stress that in our EFT approach these processes
involving different numbers of nucleons
can be treated on the same footing.
A concise account of the present study
was previously given in \cite{PMetal2001}
for the $pp$ process,
and in \cite{PMetal2} for the $hep$ process.

The reactions (\ref{pp}) and (\ref{hep}) figure
importantly in astrophysics and particle physics;
they have much
bearing upon issues of great current interest such as, for example, the solar
neutrino problem and non-standard physics in the neutrino sector.
Since the thermal energy of the interior of the Sun
is of the order of keV, and since no experimental data is
available for such low-energy regimes, one must rely on theory for
determining the astrophysical $S$-factors of the solar nuclear
processes. Here we concentrate on the threshold $S$-factor,
$S(0)$, for the reactions (\ref{pp}) and (\ref{hep}). The
necessity of a very accurate estimate of the threshold $S$-factor
for the $pp$ process, $S_{pp}(0)$, comes from the fact that $pp$
fusion essentially governs the solar burning rate and the vast
majority of the solar neutrinos come from this reaction.
Meanwhile, the $hep$ process is important in a different context.
The $hep$ reaction can produce the highest-energy solar neutrinos with
their spectrum extending beyond the maximum energy of the
${}^8\mbox{B}$ neutrinos. Therefore, even though the flux of the
$hep$ neutrinos is small, there can be, at some level, a
significant distortion of the higher end of the ${}^8\mbox{B}$
neutrino spectrum due to the $hep$ neutrinos. This change can
influence the interpretation of the results of a recent
Super-Kamiokande experiment that have generated many controversies
related to neutrino oscillations~\cite{controversy,monderen}.
To address these issues quantitatively, a reliable estimate
of $S_{hep}(0)$ is indispensable.

The primary amplitudes for both 
the $pp$ and $hep$ processes
are of the Gamow-Teller (GT) type
($\Delta J=1$, no parity change).
Since the single-particle GT operator is well known
at low energy,
a major theoretical task is the accurate estimation
of the meson-exchange current (MEC) contributions.

The nature of the specific challenge involved here can be
elucidated in terms of the {\it chiral filter} picture. If the MEC
in a given transition receives an unsuppressed contribution from a
one-soft-pion exchange diagram, then we can take advantage of the
fact that the soft-pion-exchange MEC is uniquely dictated by
chiral symmetry~\footnote{A more modern and complete
discussion on this observation has recently been given by Ananyan,
Serot and Walecka~\cite{ananyan}.} and that there is a mechanism
(called the chiral filter mechanism) that suppresses higher
chiral-order terms~\cite{KDR,BR2001}. We refer to a transition
amplitude to which the chiral filter mechanism is applicable (not
applicable) as a chiral-protected (chiral-unprotected) case. It is
known that the space component of the vector current and the time
component of the axial current are chiral-protected, whereas the
time component of the vector current and the space component of
the axial current are chiral-unprotected (see Appendix A). This
implies among other things that the isovector M1 and axial-charge
transitions are chiral-protected~\cite{M1,axialch}, but that the
GT transition is chiral-unprotected. This feature renders the
estimation of the GT amplitude a more subtle problem; the physics
behind it is that MEC here receives significant short-ranged
contributions the strength of which cannot be determined by chiral
symmetry alone.

The difficulty becomes particularly pronounced for the
$hep$ process for the following reasons.  First, the
single-particle GT matrix element for the $hep$ process is
strongly suppressed due to the symmetries of the initial and final
state wave functions. Secondly, as pointed out in
Refs.~\cite{CRSW91} (referred to as ``CRSW91") and \cite{SWPC92}
(referred to as ``SWPC92"), the main two-body corrections to the
``leading" one-body GT term tend to come with the opposite sign
causing a large cancellation. A recent detailed SNPA calculation
by Marcucci {\etal}~\cite{MSVKRB}, hereafter referred to as MSVKRB,
has re-confirmed the substantial
cancellation between the one-body and two-body terms for the $hep$
GT transition. The two-body terms therefore need to be calculated
with great precision, which is a highly non-trivial task. Indeed,
an accurate evaluation of the $hep$ rate has been a long-standing
challenge in nuclear physics~\cite{challenge}. The degree of this
difficulty may be appreciated by noting that theoretical estimates
of the $hep$ $S$-factor have varied by orders of magnitude in the
literature.

 As mentioned, in obtaining accurate estimates of the GT
transition amplitudes, it is imperative to have good theoretical
control of short-distance physics. We expect that a
``first-principle" approach based on effective field theory (EFT)
will provide a valuable insight into this issue.  We therefore
adopt here the approach developed in Refs.~\cite{BR2001,PKMR},
which purports to combine the highly sophisticated SNPA with an
EFT based on chiral dynamics of QCD. Our starting point is the
observation that, to high accuracy, the leading-order
single-particle operators in SNPA and EFT are identical, and that
their matrix elements can be reliably estimated with the use of
realistic SNPA wave functions for the initial and final nuclear
states. Next, we note that in EFT the operators representing
two-body corrections\footnote{The argument made here should apply
generally to $n$-body currents ($n\geq 2$) but since the 2-body
terms are dominant, we shall continue to restrict our discussion
to the latter.} to the leading-order one-body term can be
controlled by systematic chiral expansion in heavy-baryon chiral
perturbation theory (HB$\chi$PT)~\cite{PKMR}. Then, since the
ratio of a two-body matrix element to the leading-order one-body
matrix element can be evaluated with sufficient accuracy with the
use of the realistic SNPA wave functions\footnote{This statement
holds only for the finite-range part of two-body operators, with
the zero-range part requiring a regularization to be specified
below.}, we are in a position to obtain a reliable estimate of the
total (one-body + two-body) contribution. This approach takes full
advantage of the extreme high accuracy of the wave functions
achieved in SNPA while securing a good control of the transition
operators via systematic chiral expansion. For convenience, we
will refer to this method, which exploits the powers of {\em both}
SNPA and EFT, as ``MEEFT'' (short for {\em more effective} EFT).
MEEFT -- which is close in spirit to Weinberg's original
scheme~\cite{weinberg} based on the chiral expansion of
``irreducible terms" -- has been found 
to have an amazing predictive
power for the $n + p\rightarrow d+\gamma$ process~\cite{M1,npp}
and several other processes~\cite{beane}.
An alternative approach, 
which however is in line with our reasoning,
has been discussed 
by Ananyan, Serot and Walecka~\cite{ananyan}.

 An early HB$\chi$PT study of the $pp$ process was
made in
Ref.~\cite{pp} (hereafter referred to as PKMR98)
by four of the authors.
The calculation in PKMR98 was carried out up to
next-to-next-to-next-to-leading order (\nlo3) in chiral counting
(see below). At \nlo3, two-body meson-exchange currents (MEC)
begin to contribute, and there appears one unknown parameter in
the chiral Lagrangian contributing to the MEC. This unknown
constant, called $\dR$ in Ref.~\cite{pp}, represents the strength of a
four-nucleon-axial-current contact interaction. In Ref.~\cite{pp},
since no method was known to fix the value of $\dR$, the
$\dR$-term was simply ignored by invoking a qualitative argument
that the short-range repulsive core would strongly suppress its
contribution. Due to uncertainties associated with this argument,
Ref.~\cite{pp} was unable to corroborate or exclude the result of
the latest SNPA calculation \cite{TBDexp}, $\delta_{\rm 2B}=0.5
\sim 0.8$ \%, where $\delta_{\rm 2B}$ is the ratio of the
contribution of the two-body MEC to that of the one-body current
(see below).

The situation can be greatly improved by using MEEFT.
As first discussed in Refs.\cite{PMetal2001,PMetal2}
and as will be expounded here,
the crucial point is that exactly the same combination of counter
terms that defines the constant $\dR$ enters into the Gamow-Teller
(GT) matrix elements that feature in $pp$ fusion, tritium
$\beta$-decay, the $hep$ process, $\mu$-capture on a deuteron, and
$\nu$--$d$ scattering and that the short-range interaction
involving
the constant $\dR$ is expected to be ``universal,"
that is, $A$-independent. Therefore, assuming that three- and
four-body currents can be ignored (which we will justify a
posteriori), if the value of $\dR$ can be fixed using one of
the above
processes, then we can make a totally parameter-free prediction
for the GT matrix elements of the other processes.  Indeed,
the existence of accurate experimental data for the tritium
$\beta$-decay rate, $\Gamma_\beta^t$, and the availability of
extremely well tested realistic wave functions for the $A$=3
nuclear systems allow us to carry out this program. In the present
work we determine the value of $\dR$ from $\Gamma_\beta^t$ and
perform parameter-free EFT-based calculations of $S_{pp}(0)$ and
$S_{hep}(0)$.

 As described below, our scheme has a cutoff parameter
$\Lambda$, which defines the energy/momentum cutoff scale of EFT
below which reside the chosen explicit degrees
of freedom\footnote{The
cutoff specifies not just the relevant degrees of freedom but also
their momentum/energy content. This should be understood in what
follows although we do not always mention it.}. Obviously, in
order for our result to be physically acceptable, its cutoff
dependence must be under control. In our scheme, for a given value
of $\Lambda$ in a physically reasonable range (to be discussed
later), $\dR$ is determined to reproduce $\Gamma_\beta^t$; thus
$\dR$ is a function of $\Lambda$. According to the premise
of EFT, even if $\dR$ itself is $\Lambda$-dependent, physical
observables (in our case the $S$-factors) should be independent of
$\Lambda$ as required by renormalization-group invariance. We
shall show that our results meet this requirement to a
satisfactory degree. The robustness of our calculational results
against changes in $\Lambda$ allows us to make predictions on
$S_{pp}(0)$ and $S_{hep}(0)$ with much higher precision than
hitherto achieved. Thus we predict: $S_{pp}(0) = 3.94\!\times\!(1
\pm 0.004) \!\times\!10^{-25}\ \mbox{MeV-b}$ and $S_{hep}(0) =
(8.6 \pm 1.3)\!\times\! 10^{-20} \ \mbox{keV-b}$.

The remainder of this article is organized as follows. In Section
II we briefly explain our formalism; in particular, we describe
the relevant transition operators derived in HB$\chi$PT. Section
III presents the calculation of $S_{pp}(0)$, while Section IV is
concerned with the estimation of $S_{hep}(0)$. Section V is
devoted to discussion and conclusions. We have made the
explanation of the formalism
in the text as brief and focused as possible,
relegating most technical details to the Appendices.

\section{Formalism}
 We sketch here the basic elements of our formalism.
The explicit degrees of freedom
taken into account in our scheme are
the nucleon and the pion, with all other degrees of freedom
($\rho$- and $\omega$-mesons, $\Delta (1232)$, etc.) integrated
out.
The HB$\chi$PT Lagrangian can be written as
\be
{\cal L} =
\sum_\lambda{\cal L}_\lambda =
{\cal L}_0 + {\cal L}_1 + \cdots,
\ee
with the chiral order $\lambda$
defined as
\be \lambda \equiv
d+e+\frac{n}{2}-2 \ ,
\label{lambda}
\ee
where $d$, $e$ and $n$ are, respectively,
the numbers of derivatives
(the pion mass counted as one derivative),
external fields
and nucleon lines belonging to a vertex.
Chiral symmetry requires $\lambda\ge 0$.
The leading-order Lagrangian is given by
\be {\cal L}_0 &=& {\bar B}\left[ i v\cdot
D + 2 i g_A S\cdot \Delta \right] B
- \frac{1}{2} \sum_A C_A \left({\bar B} \Gamma_A B\right)^2
\nonumber\\
&+& f_\pi^2 {\rm Tr}\left(i \Delta^\mu i \Delta_\mu\right) +
\frac{f_\pi^2}{4} {\rm Tr}(\chi_+)  \ , \label{chiralag2}
 \ee
where $B$ is the nucleon field in HB$\chi$PT;
$g_A= 1.2670 \pm 0.0035$ 
is the axial-vector coupling constant\cite{PDG}, 
and $f_\pi=92.4$ MeV is the pion decay constant. 
Furthermore
 \be
D_\mu B &=& (\del_\mu + \Gamma_\mu) B ,
\nonumber \\
\Gamma_\mu &=& \frac{1}{2} \left[\xi^\dagger,\, \del_\mu
\xi\right] -\frac{i}{2}\xi^\dagger R_\mu \xi - \frac{i}{2}
\xi L_\mu\xi^\dagger,
\nonumber \\
\Delta_\mu &=& \frac{1}{2} \left[\xi^\dagger,\, \del_\mu
\xi\right] +\frac{i}{2}\xi^\dagger R_\mu \xi - \frac{i}{2}
\xi L_\mu\xi^\dagger,
\nonumber \\
\chi_+ &=& \xi^\dagger \chi \xi^\dagger + \xi \chi^\dagger \xi \ ,
\label{deltamu}\ee
with
 \be \xi = \sqrt{\Sigma} =
  {\rm exp}\left(i\frac{{\vtau}\cdot
  {\vec \pi}}{2 f_\pi}\right).
\ee
$R_\mu \equiv \frac{\tau^a}{2} \left(
  {\cal V}^a_\mu + {\cal A}^a_\nu\right)$
and
$L_\mu = \frac{\tau^a}{2} \left(
  {\cal V}^a_\mu - {\cal A}^a_\nu\right)$
denote external gauge fields, and $\chi$ is proportional to the
quark mass matrix. If we neglect the small isospin-symmetry
breaking, then
$\chi=m_\pi^2$ (in the absence of external
scalar and pseudo-scalar fields). For convenience, we work in the
reference frame in which the four-velocity $v^{\mu}$ and the spin
operator $S^{\mu}$ are \be v^\mu = (1,\, \vzero) \ \ \
\mbox{and} \ \ S^\mu = \left(0,\, \frac{\vs}{2}\right).
\ee The NLO Lagrangian (the so-called ``$1/m$" term) in the
one-nucleon sector is given in  Ref.~\cite{bernard}, while that in the
two-nucleon sector is given in Ref.~\cite{Kolck3}\footnote{ Our
definition of the pion field here is different from that used in
Ref. \cite{Kolck3}; we have changed the sign of the pion field.
Furthermore, we employ here manifestly Lorentz-invariant and
chiral-invariant interactions.}. With four-fermion contact terms
included, the Lagrangian takes the form
 \be
{\cal L}_1 &=& {\bar B} \left\{
  \frac{v^\mu v^\nu - g^{\mu\nu}}{2 m_N} D_\mu D_\nu
   + 4 c_3 i\Delta\cdot i\Delta\right.\nonumber\\
   &&\left. + \left(2 c_4 + \frac{1}{2m_N}\right)
  \left[S^\mu, \,S^\nu\right] \left[ i \Delta_\nu,\, i\Delta_\nu\right]
\right.
\nonumber \\
&&\left.
   -\ i \frac{1+c_6}{m_N}
  \left[S^\mu, \,S^\nu\right] f_{\mu\nu}^+ \right\}B
 -\ 4 i d_1 \,
 {\bar B} S\cdot \Delta B\, {\bar B} B\nonumber\\
&& + 2 i d_2 \,
  \epsilon^{abc}\,\epsilon_{\mu\nu\lambda\delta} v^\mu \Delta^{\nu,a}
 {\bar B} S^\lambda \tau^b B\, {\bar B} S^\delta \tau^c
 B\nonumber\\
&&+ \cdots, \label{Lag1} \ee
where
$m_N \simeq$ 939 MeV is the nucleon mass, and
\be
f_{\mu\nu}^+ &=&
 \xi (\del_\mu L_\nu - \del_\nu L_\mu
   - i \left[L_\mu,\,L_\nu\right]) \xi^\dagger
   \nonumber \\
 && \;\;\;\;+
 \xi^\dagger (\del_\mu R_\nu - \del_\nu R_\mu
   - i \left[R_\mu,\,R_\nu\right]) \xi,
\ee $\epsilon_{0123}=1$, and $\Delta_\mu = \frac{\tau^a}{2}
\Delta^a_\mu$. We have shown here only those terms which are
directly relevant to our present study. The dimensionless
low-energy-constants (LECs),
$\hat c$'s and $\hat d$'s, are defined as \be c_{3,4} =
\frac{1}{m_N}\, \hat c_{3,4}, \ \ \ d_{1,2} = \frac{g_A}{m_N
f_\pi^2}\,\hat d_{1,2}. \ee

\vspace*{0.6cm}
We now consider the chiral counting of the electroweak currents
(see the Appendices for details).
In the present scheme it is sufficient to focus on
``irreducible graphs" in Weinberg's
classification~\cite{weinberg}. 
Irreducible graphs are organized
according the chiral index $\nu$ given by
 \be \nu = 2 (A-C) + 2 L
+\sum_i \nu_i\,, \label{nu}
 \ee
where $A$ is the number of nucleons involved in the process, $C$
the number of disconnected parts, and $L$ the number of loops;
$\nu_i$ is the chiral index $\lambda$, eq.(\ref{lambda}),
of the $i$-th vertex.
One can show that a diagram 
characterized by eq.(\ref{nu})
involves an $n_B$-body transition operator, 
where $n_B \equiv A- C+1$.
The physical amplitude is
expanded with respect to $\nu$.
As explained at length in the Appendix,
the leading-order one-body GT
operator belongs to $\nu$=0. Compared with this operator, a
Feynman diagram with a chiral index $\nu$ is suppressed by a
factor of $(Q/\Lambda_\chi)^\nu$, where $Q$ is a typical
three-momentum scale or the pion mass,
and $\Lambda_\chi \sim$ 1 GeV is
the chiral scale.\footnote{ For convenience, a chiral order
corresponding to $\nu$ is often referred to as \nlo{\nu}; $\nu$=1
corresponds to NLO (next-to-leading order), $\nu$=2 to N$^2$LO
(next-to-next-to-leading order), and so on.}
In our case it is important to take into account the kinematic
suppression of the time component of the nucleon four-momentum.
We note \be v\cdot p_l \sim v\cdot p_l' \sim v\cdot k_l \sim
\frac{Q^2}{m_N}, \label{kinetic} \ee where $p_l^\mu$ ($p_l'^\mu$)
denotes the initial (final) momentum of the $l$-th nucleon, and
$k_l^\mu \equiv (p_l'-p_l)^\mu$.
Therefore, each appearance of
$v\cdot p_l$, $v\cdot p_l'$ or $v\cdot k_l $ carries two powers of
$Q$ instead of one, 
which implies that $\nu$ increases by two
units rather than one.
It is also to be noted 
that, if we denote by $q^\mu=(q_0,\vq)$
the momentum transferred to the leptonic pair
in eqs.(\ref{pp}) (\ref{hep}), 
then $q_0\sim |\vq|$$\sim Q^2/\Lambda_\chi$
$\sim{\cal O}(Q^2)$
rather than ${\cal O}(Q)$
as naive counting would suggest.
These features turn out to 
simplify our calculation considerably. 

In this paper, as far as the main calculation
is concerned, we shall limit ourselves to \nlo3;
for certain discussions, however,
we shall consider operators
belonging to \nlo4 as well.

We now describe the derivation of one-body (1B) and
two-body (2B) current operators 
with due consideration of chiral counting. 
The current in momentum space is written as
 \be J^\mu(\vq)= V^\mu(\vq) + A^\mu(\vq)
 = \int\! d\vx\, \e^{- i\vq\cdot \vx} J^\mu(\vx).
\ee
When necessity arises to distinguish
the space and time components of the currents,
we use the notations
\be
V^\mu=(V^0,\, \vV)\,,\;\;\;\; A^\mu=(A^0,\,\vA).
\ee

For the clarity of presentation, 
we first give a summary chart 
of the basic chiral counting characteristics of
the relevant currents,
and then provide more detailed explanations
in the remainder of this section
and in the Appendices. 
The chiral counting of the electroweak currents is
summarized in Table \ref{power}, 
where the non-vanishing contributions
at $\vq=0$ are indicated.\footnote{
For small but finite values of $\vq\ne 0$, 
there are slight deviations
from this chart; for instance,
for all the four cases in the table,
there arise one-body contributions at
NLO and at higher orders.
However, these deviations are not significant
in our case, which involves very small values
of $\vq$.}
 
We now discuss the entries 
of this table order by order:

\begin{table}[htb]
\caption{\protect Contributions from each type of current at
$\vq=0$. The entry of ``$-$" indicates the absence of
contribution. ``1B-RC'' stands for relativistic corrections to the
one-body operators, and ``2B-1L'' for one-loop 2-body
contributions including counter term contributions. \label{power}}
\begin{tabular}{|c||l|l|l|l|l|}
$J^\mu$ & LO & NLO & \nlo2 & \nlo3 & \nlo4
\\ \hline
$\vA$ & 1B & $-$ & 1B-RC & 2B & 1B-RC, 2B-1L and 3B \\
$A^0$ & $-$ & 1B & 2B & 1B-RC & 1B-RC, 2B-1L \\
$\vV$ & $-$ & 1B & 2B & 1B-RC & 1B-RC, 2B-1L \\
$V^0$ & 1B & $-$ & $-$  & 2B & 1B-RC, 2B-1L and 3B \\
\end{tabular}
\end{table}

\begin{itemize}
\item
$\nu$ = 0 : One-body $\vA$ and $V^0$: $\vA$ gives the Gamow-Teller
(GT) operator, while $V^0$ is responsible for the charge operator.

\item
$\nu$ = 1 : One-body $A^0$ and $\vV$: $A^0$ gives the
axial-charge operator while $\vV$ gives the M1 operator.
\item
$\nu$ = 2 : Two-body tree current with $\nu_i=0$ vertices, namely,
the soft-pion-exchange current. 
This is a leading correction to
the one-body M1 and axial-charge operators 
carrying an odd orbital
angular momentum.
\item
$\nu$ = 3 : Two-body tree currents with $\sum_i \nu_i=1$, which
correspond to the hard-pion current, considered in
CRSW91~\cite{CRSW91} and SWPC92~\cite{SWPC92}. These are leading
corrections to the GT and $V^0$ operators carrying an even orbital
angular momentum.

\item
$\nu$ = 4 : All the components of the electroweak current receive
contributions of this order. They consist of two-body one-loop
corrections as well as leading-order (tree) three-body
corrections. Among the three-body currents, however, there are no
six-fermion contact terms proportional to $(\bar N N)^3$, because
there is no derivative at the vertex and hence no external field.
\end{itemize}

It is noteworthy 
that the counting rule for $\vV$ is the same
as for $A^0$, and the counting rules for $V^0$ and $\vA$ are the
same.  
The behavior of $\vV$ and $A^0$ summarized in
Table~\ref{power} represents the chiral filter
mechanism~\cite{KDR}, and $\vV$ and $A^0$ are referred to as
chiral-filter-protected currents.
By contrast, $V^0$ and $\vA$
belong to chiral-filter-unprotected currents.

We now discuss the explicit expressions
for the relevant currents.
For the one-body (1B) currents,
for both the vector and axial cases, 
one can simply carry over the expressions
obtained in MSVKRB.
Up to \nlo3, the 1B currents 
in coordinate representaion are given as 
\be 
\tilde V^0(l)
&=& \tau_l^- \e^{-i \vq\cdot \vr_l} \left[
 1 + i \vq \cdot \vs_l \times \vp_l
\frac{2 \mu_V -1}{4 m_N^2}
 \right],
\nonumber \\
\tilde \vV(l) &=& \tau_l^-  \e^{-i \vq\cdot \vr_l} \left[
 \frac{\vbp_l}{m_N}\left(1 - \frac{\vbp_l^2}{2 m_N^2}\right)
 + i \frac{\mu_V}{2 m_N} \vq \times \vs_l\right.\nonumber\\
 &&\left. + i \vs_l\times \vbp_l\, q_0 \frac{2 \mu_V-1}{4 m_N^2}
 \right],
\nonumber \\
\tilde A^0(l) &=& - g_A \tau_l^- \e^{-i \vq\cdot \vr_l} \left[
 \frac{\vs_l\cdot \vbp_l}{m_N}
 \left(1 - \frac{\vbp_l^2}{2 m_N^2}\right)
 \right],
\nonumber \\
\tilde \vA(l) &=& - g_A \tau_l^- \e^{-i \vq\cdot \vr_l} \left[
 \vs_l \right.\nonumber\\
 &&\left. + \frac{2 (\vbp_l\, \vs_l \cdot \vbp_l - \vs_l \, \vbp_l^2)
     + i \vq\times \vbp_l}{4 m_N^2}
 \right],
 \label{J1Bnon}\ee
where $\mu_V \simeq 4.70$ is the isovector
anomalous magnetic moment of the nucleon.
Eq.(\ref{J1Bnon}) gives the isospin-lowering
currents, 
\be J_\mu \equiv J_\mu^{a=1} - i J_\mu^{a=2}, \ee 
and 
$\tau_l^- \equiv 
\frac12 (\tau_l^x - i\tau_l^y)$. 
The tildes in eq.(\ref{J1Bnon}) imply 
that the currents are given 
in the coordinate space representation,
and $\vp_l = -i \nabla_l$ and $\vbp_l =
-\frac{i}{2} \left( \stackrel{\rightarrow}{\nabla}_l -
\stackrel{\leftarrow}{\nabla}_l\right)$ 
act on the wave functions.

We next discuss the two-body (2B) currents.
The expressions for the $\vV_{\rm 2B}$
and $A^0_{\rm 2B}$ operators
can be found in \cite{PKMR,PJM}.
The $V^0_{\rm 2B}$ operator does not appear
up to the order under consideration.
The derivation of the 2B axial current,
$\vA_{\rm 2B}$, in HB$\chi$PT  
is described in Appendix A.
In momentum space, $\vA_{\rm 2B}$ reads
\be
\vA_{\rm 2B} &=& \sum_{l<m}^{A} \vA_{lm} ,
\nonumber\\
\vA_{12} &=&
\frac{g_A}{2 m_N f_\pi^2}\, \frac{1}{m_\pi^2 + \vk^2}
 \Bigg[ - \frac{i}{2} \tau^-_\times\,\vp\,\,
     (\vs_1-\vs_2)\cdot\vk
 \nonumber \\
 &&+ 4 \,\hat c_3\, \vk \,\, \vk\cdot
  (\tau_1^- \vs_1 +\tau_2^- \vs_2)
\nonumber \\
 &&+ \left(\hat c_4
  + \frac14\right) \tau^-_\times\,
 \vk \times \left[ \vs_\times \times \vk\, \right]
 \frac{}{}\Bigg]
\nonumber \\
 &&- \frac{g_A}{m_N f_\pi^2} \left[
  2 \hat d_1 (\tau_1^- \vs_1 + \tau_2^- \vs_2)
 + \hat d_2 \tau^a_\times \vs_\times
 \right] \>\>,\label{2-body}
\ee
with $\vp \equiv (\vbp_1 - \vbp_2)/2$,
$\vbp_l\equiv (\vp_l+ \vp_l^{\,\prime})/2$,
$\tau_l^- \equiv \frac12 (\tau_l^x - i \tau_l^y)$,
$\tau^a_\times\equiv (\tau_1\times\tau_2)^x
 - i (\tau_1\times\tau_2)^y$,
and similarly for $\vs_\times$;
$\hat c$'s and $\hat d$'s are the LECs
explained in PKMR98.
The values of $\hat c$'s in Eq.(\ref{2-body})
have been determined from $\pi$-$N$ data \cite{csTREE}:
${\hat c}_3= -3.66 \pm 0.08$
and ${\hat c}_4= 2.11 \pm 0.08$.
The two constants,
$\hat d_{1}$ and $\hat d_2$, remain to be fixed
but it turns out (see Appendix C.2)
that, thanks to Fermi-Dirac statistics,
only one combination of them,
\be
 \dR\equiv \hat d_1 +2 \hat d_2 +
\frac13 \hat c_3
 + \frac23 \hat c_4 + \frac16 \>\>
\label{dr}
\ee
is relevant in the present context.\footnote{
A sign error made
in the expression for $\dR$ in PKMR98
has been corrected here.}

It should be noted that the two-body currents
given in eqs.~(\ref{2-body})
are valid only up to a certain cutoff $\Lambda$.
This implies that, when we go to coordinate space,
the currents must be regulated.
This is a key point in our approach.
Specifically, in performing Fourier transformation
to derive the $r$-space representation of
a transition operator,
we use the Gaussian regularization
(see Appendix C).
This is, to good accuracy, equivalent
to replacing the delta and Yukawa functions
with the corresponding regulated functions,
\be
\delta_\Lambda^{(3)}(r)&\equiv&
 \int \!\!\frac{d^3 \vk}{(2\pi)^3}\,
  S_\Lambda^2(\vk^2)\, \e^{ i \vk\cdot \vr}\,,
 \nonumber\\
  y_{0\Lambda}^\pi(r)&\equiv&
  \int \!\!\frac{d^3 \vk}{(2\pi)^3}\,
  S_\Lambda^2(\vk^2)\, \e^{ i \vk\cdot \vr}
  \frac{1}{\vk^2 + m_\pi^2}
  \nonumber\\
y_{1\Lambda}^\pi(r) &\equiv&
- r \frac{\del}{\del r} y_{0\Lambda}^\pi(r)
  \nonumber\\
y_{2\Lambda}^\pi(r) &\equiv &
\frac{1}{m_\pi^2}
  r \frac{\del}{\del r} \frac{1}{r}
  \frac{\del}{\del r} y_{0\Lambda}^\pi(r)\,,
\ee
where the cut-off function, $S_\Lambda(\vk^2)$,
is defined as
\be
S_\Lambda(\vk^2) =
\exp\!\left(\!-\frac{\vk^2}{2\Lambda^2}\!\right).
\label{regulator}
\ee
The resulting $r$-space expressions of the currents
in the center-of-mass (c.m.) frame are
\be
\vV_{12}(\vr) &=& -
\frac{g_A^2 m_\pi^2}{12 f_\pi^2}
\tau_\times^- \,
 \vr \,
 \left[
 \vs_1\cdot\vs_2 \, y_{0\Lambda}^\pi(r)
 + S_{12} \, y_{2\Lambda}^\pi(r) \right]
\nonumber \\
 &-& i \frac{g_A^2}{8 f_\pi^2}
 \vq\times \left[
 \voO_\times y_{0\Lambda}^\pi(r)
 +\left( \voT_\times - \frac23 \voO_\times \right)
  y_{1\Lambda}^\pi(r)
 \right],
\nonumber \\
A^{0}_{12}(\vr) &=& -
\frac{g_A}{4 f_\pi^2}
\tau_\times^-
\left[
\frac{\vs_+ \cdot \hatr}{r}
+
\frac{i}{2} \vq\cdot \hatr\, \vs_-\cdot \hatr\,
\right]
y_{1\Lambda}^\pi(r) ,
\nonumber \\
\vA_{12}(\vr) &=&
- \frac{g_A m_\pi^2}{2 m_N f_\pi^2}
\Bigg[
\nonumber \\
&&\left[
\frac{\hat c_3}{3} (\voO_+ + \voO_-)
+\frac23 \left(\hat c_4 + \frac14\right)
   \voO_\times \right] y_{0\Lambda}^\pi(r)
\nonumber \\
&&
  + \left[
     \hat c_3 (\voT_+ + \voT_-)
  - \left(\hat c_4 + \frac14\right) \voT_\times
     \right] y_{2\Lambda}^\pi(r)
   \Bigg]
\nonumber \\
&+& \frac{g_A}{2 m_N f_\pi^2 }
\Big[
\frac{1}{2} \tau_\times^-
   (\vbp_1 \,\vs_2\cdot\hatr +
   \vbp_2\,\vs_1\cdot\hatr)
\frac{y_{1\Lambda}^\pi(r)}{r}
\nonumber \\
&&
 + \delta_\Lambda(r)
 \, \hat d^R \voO_\times
\Big],
\label{vAnuFT}\ee
where $\vr=\vr_1 - \vr_2$,
$S_{12}= 3 \vs_1\cdot \hatr\,\vs_2\cdot \hatr - \vs_1\cdot\vs_2$,
and
$\voO_{\odot}^{k} \equiv \tau_\odot^- \sigma_\odot^k$,
$ \voO_{\odot} \equiv \tau_\odot^- \vs_\odot$,
$ \voT_{\odot} \equiv
 \hatr\, \hatr\cdot \voO_{\odot} - \frac13 \voO_{\odot}$,
$\odot=\pm,\times$,
$\tau_\odot^-\equiv(\tau_1\odot\tau_2)^-\equiv
(\tau_1\odot\tau_2)^x -i (\tau_1\odot\tau_2)^y$ and
$\vs_\odot\equiv(\vs_1\odot\vs_2)$.
 We emphasize again
that $\vA_{12}$ in Eq.(\ref{vAnuFT})
contains only one unknown LEC, $\dR$,
that needs to be fixed using an empirical input.
As mentioned in Section I,
we choose here to determine $\dR$ using
the experimental value of $\Gamma_\beta^t$.

\section{Determination of $\dR$}

 The cutoff parameter $\Lambda$ characterizes the
energy-momentum scale of our EFT. A reasonable range of $\Lambda$
may be inferred as follows. According to the general {\it tenet}
of $\chi$PT, $\Lambda$ larger than $\Lambda_\chi \simeq 4\pi f_\pi
\simeq m_N$ has no physical meaning. Meanwhile, since the pion is
an explicit degree of freedom in our scheme,
$\Lambda$ should be much larger
than the pion mass to ascertain that genuine low-energy
contributions are properly included. These considerations lead us
to adopt $\Lambda$ = 500-800 MeV as a natural
range.

In the present work
we use as representative values
$\Lambda$ = 500, 600 and 800 MeV,
and for each of these values of $\Lambda$
we adjust $\dR$
to reproduce the experimental value of
$\Gamma_\beta^t$.
With the use of the value of $\dR$
so determined,
we evaluate the $pp$ and the $hep$ amplitudes.\footnote{
The masses of the light-quark vector mesons
($\rho$ and $\omega$) are less than 800 MeV;
therefore, with the use of $\Lambda=800$ MeV,
an accurate description of certain observables
might require the explicit presence
of the vector mesons, even though
the probe energy/momentum in question
is much smaller than 800 MeV.
Thus the results with the cutoff $\Lambda=800$ MeV should be
taken with some caution.}

To determine $\dR$ from $\Gamma_\beta^t$,
we calculate $\Gamma_\beta^t$
from the matrix elements of the current operators
evaluated for accurate $A$=3 nuclear wave functions.
We employ here the wave functions
obtained in Refs.~\cite{MSVKRB,roccoetal}
using the correlated-hyperspherical-harmonics
(CHH) method~\cite{VKR95,VRK98}.
It is obviously important
to maintain consistency between
the treatments of the $A$=2, 3 and 4 systems.
We shall use here the same Argonne $v_{18}$
(AV18) potential~\cite{av18}
for all these nuclei.
For the $A\ge 3$ systems we add
the Urbana-IX (AV18/UIX) three-nucleon potential~\cite{uix}.
Furthermore, we apply the same regularization method
to all the systems
in order to control short-range physics
in a consistent manner.

The values of $\dR$ determined in this manner are:
\be
\dR &=& 1.00\pm0.07\;\;\;\;\;\;\;
{\rm for}\;\;\Lambda=500\; {\rm MeV}\,,
\nonumber\\
\dR &=& 1.78\pm0.08\;\;\;\;\;\;\;
{\rm for}\;\;\Lambda=600\; {\rm MeV}\,,
\label{dR}\\
\dR &= &3.10\pm0.10\;\;\;\;\;\;\;
{\rm for}\;\;\Lambda=800\; {\rm MeV}\,,
\nonumber
\ee
where the errors correspond to
the experimental uncertainty
in $\Gamma_\beta^t$.
Once $\dR$ has been determined,
we are in a position to make
a parameter-free calculation of
the transition amplitudes
for $pp$ and $hep$,
which will be described
in the next two sections.

\section{The \lowercase{$pp$} process}

It is convenient to decompose the matrix element of the GT
operator into one-body and two-body parts
 \be \calM= \calM_{\rm
1B} + \calM_{\rm 2B} \>\>. \ee We discuss them separately. In
PKMR98, an extensive analysis was made of the leading-order (LO)
one-body matrix element $\calM_{\rm 1B}^{C+N}$, with a focus on
the connection between EFT and the effective range expansion. The
results obtained with the AV18 potential~\cite{av18} were
 \be \calM_{\rm 1B}^{C+N} &=& ( 1 \mp 0.02\ \% \mp 0.07\
\% \mp 0.02\ \%)\nonumber\\
 && \times 4.859\ \fm \label{m1b} \>\>,
\ee where the errors are due to uncertainties in the scattering
length and effective ranges. The ``full" one-body contribution in
PKMR98 includes the vacuum-polarization (VP) and
two-photon-exchange (C2) contributions. In PKMR98, however, the
one-body current due to the $1/m_N^2$ term in eq.(\ref{A1Bnon})
was ignored. Although this term is required for formal
consistency, its numerical value turns out to be quite small,
$\calM_{\rm 1B}^{1/m_N^2} = -0.006\ \fm$. In terms of the
$\Lambda_{pp}$ defined in Ref.~\cite{KB94},\footnote{The subscript
$pp$ has been added here to avoid confusion with the cutoff
parameter $\Lambda$.
} we have
\be
\Lambda_{pp}^2&\equiv& \frac{|a^C|^2 \gamma^3}{2}\, A_S^2\, \calM_{\rm
1B}^2
= 6.91 \ee
for the central value, where $a^C$ is the $pp$ $^1$S$_0$
scattering length, and $\gamma$ and $A_S$ are the wave number and
S-wave normalization constant pertinent to the deuteron,
respectively. This should be compared with $6.93$ obtained in
Ref.~\cite{pp}.

The properly regularized two-body GT matrix
elements for the $pp$ process read
\be \calM_{\rm 2B} &=&
 \frac{2}{m_N f_\pi^2}
 \int_0^\infty\! dr\,  \left\{ \frac{}{} \right.
\nonumber \\
&&
 \frac{m_\pi^2}{3}
\left(\hat c_3 + 2 \hat c_4 + \frac12\right)
  y_{0\Lambda}^\pi(r)\, u_d(r)\,  u_{pp}(r)
\nonumber\\
&-& \sqrt{2}
\frac{m_\pi^2}{3}
 \left(\hat c_3 - \hat c_4 - \frac14\right)
 y_{2\Lambda}^\pi(r)\, w_d(r) \, u_{pp}(r)
\nonumber \\
&+&
 \frac{y_{1\Lambda}^\pi(r)}{12 r} \Bigg[
 \left[ u_d(r)-\sqrt2 w_d(r)\right]
 u_{pp}'(r) \nonumber \\
&-& \left[ u_d'(r)-\sqrt2 w_d'(r)\right] u_{pp}(r)
+ \frac{3\sqrt2}{r} w_d(r) u_{pp}(r)
\Bigg] \nonumber \\
&-&\left. \dR
\delta_\Lambda^{(3)}(r)\, u_d(r) u_{pp}(r)
\frac{}{}\right\},
\label{calM2Bdelta}
\ee
where $u_d(r)$ and $w_d(r)$ are
the S- and D-wave components
of the deuteron wave function, and $u_{pp}(r)$
is the $^1$S$_0$ $pp$ scattering
wave (at zero relative energy).
The results are given
for the three representative values
of $\Lambda$ in Table~\ref{tb:tb1};
for convenience, the values of $\dR$
given in Eq.(\ref{dR}) are also listed.
The table indicates
that, although the value of $\dR$
is sensitive to $\Lambda$,
$\calM_{\rm 2B}$ is amazingly stable
against the variation of
$\Lambda$ within the stated range.
In view of this high stability, we
believe that we are on the conservative side
in adopting the estimate
$\calM_{\rm 2B}= (0.039 \sim 0.044)\ \fm$.
Since the leading
single-particle term is independent of $\Lambda$,
the total amplitude
$\calM = \calM_{\rm 1B}+\calM_{\rm 2B}$ is
$\Lambda$-independent to the same degree
as $\calM_{\rm 2B}$.
The $\Lambda$-independence of
the physical quantity $\calM$,
which is in conformity with the {\it tenet} of EFT,
is a crucial feature
of the result in our present study.
The relative strength of the
two-body contribution as compared
with the one-body contribution is
\be
\delta_{\rm 2B} \equiv
\frac{\calM_{\rm 2B}}{\calM_{\rm 1B}}
= (0.86 \pm 0.05)\ \%.
\label{delta2B-new}
\ee
We remark that the central value of
$\delta_{\rm 2B}$ here is considerably
smaller than the corresponding value,
$\delta_{\rm 2B}= 4$ \%, in
PKMR98. Furthermore, the uncertainty of $\pm$0.05 \% in
Eq.(\ref{delta2B-new}) is drastically smaller than the
corresponding figure, $\pm$4 \%, in PKMR98.

\begin{table}[b]
\caption{\protect The strength $\dR$ of the contact term and the
two-body GT matrix element, $\calM_{\rm 2B}$, for the $pp$
process calculated for representative values of $\Lambda$.}
\begin{tabular}{|c|c|l|}
$\Lambda$ (MeV) & $\dR$ & $\calM_{\rm 2B}$ (fm) \\
\hline
500 & $1.00 \pm 0.07$ &
$0.076 - 0.035\ \dR \simeq 0.041 \pm 0.002 $\\
\hline
600 & $1.78 \pm 0.08$ &
$0.097 - 0.031\ \dR \simeq 0.042 \pm 0.002$
\\ \hline
800 & $3.90 \pm 0.10$ &
$0.129 - 0.022\ \dR \simeq 0.042 \pm 0.002$
\\
\end{tabular}
\label{tb:tb1}
\end{table}

We now turn to the threshold $S$ factor,
$S_{pp}(0)$.
Adopting the value
$G_V=(1.14939 \pm 0.00065)\times 10^{-5}\
\mbox{GeV}^{-2}$ \cite{hardy},
we obtain
\be S_{pp}(0) &=&
3.94\times
\left(\frac{1+\delta_{2B}}{1.01}\right)^2
\left(\frac{g_A}{1.2670}\right)^2
\left(\frac{\Lambda_{pp}^2}{6.91}\right)^2
\nonumber \\
&=&
 3.94\times
  (1 \pm 0.0015 \pm 0.0010 \pm \ve)
  \label{S-factor}
\ee
in units of $10^{-25}\ \mbox{MeV-b}$.
Here the first error is due to uncertainties
in the input parameters in the one-body part,
while the second error represents the uncertainties
in the two-body part;
$\ve(\approx 0.001)$ denotes possible uncertainties
due to higher chiral order contributions (see below).
To make a formally rigorous assessment of $\ve$, we must evaluate
loop corrections and higher-order counter terms.
Although an N$^4$LO calculation would not involve any new
unknown parameters, it is a non-trivial task.
Furthermore, loop corrections necessitate a more
elaborate regularization scheme since the naive cutoff
regularization used here violates chiral symmetry at loop orders.
(This difficulty, however, is not insurmountable.)
These formal
problems set aside, it seems reasonable to assess $\ve$ as
follows. 
We first recall that both tritium $\beta$-decay and solar
$pp$ fusion are dominated by the one-body GT matrix elements, the
evaluation of which is extremely well controlled from the SNPA as
well as EFT points of view.
Therefore,
the precision of our calculation is
governed by the reliability of estimation of small corrections to
the dominant one-body GT contribution. Now, we have seen that the
results of the present \nlo3 calculation nicely fit into the
picture expected from the general premise of EFT: (i) the \nlo3
contributions are indeed much smaller than the leading order term;
(ii) the physical transition amplitude $\calM$ does not depend on
the cutoff parameter.  Although these features do not constitute a
formal proof of the convergence of the chiral expansion used here, it
is {\it extremely unlikely} that higher order contributions be so
large as to completely upset the physically reasonable behavior
observed in the \nlo3 calculation. It should therefore be safe to
assign to $\ve$ an uncertainty comparable to the error estimate for
the two-body part in Eq.(\ref{S-factor}); viz., $\ve \approx
0.1$ \%.
In this connection we remark that
an axial three-body MEC contribution
to the $^3$H GT matrix element
was calculated explicitly in SNPA~\cite{MSVKRB}
and found to be negligible relative
to the leading two-body mechanisms.
This feature is consistent with the above argument
since, in the context of EFT,
the three-body MEC represents a higher-order effect
subsumed in $``\ve"$ in Eq.(\ref{S-factor}).

 Apart from the  notable numerical differences between the
present work and PKMR98, it is worth noting that short-range
physics is much better controlled in MEEFT. In the conventional
treatment of MEC, one derives the coordinate space representation
of a MEC operator by applying ordinary Fourier transformation
(with no restriction on the range of the momentum variable) to the
amplitude obtained in momentum space; this corresponds to setting
$\Lambda=\infty$ in Eq.(\ref{regulator}). In PKMR98, where this
familiar method is adopted, the $\dR$ term appears in the
zero-range form, $\dR\delta(r)$.
PKMR98 chose to introduce short-range
repulsive correlation with hard-core radius $r_C$ and eliminate
the $\dR\delta(r)$ term {\it by hand}. The remaining finite-range
terms were evaluated as functions of $r_C$. $\calM_{\rm 2B}$
calculated this way exhibited substantial $r_C$-dependence,
indicating that short-range physics was not well controlled.
Inclusion of the $\dR$ term, with its strength renormalized as
described here, properly takes into account the short-range
physics inherited from the integrated out degrees of freedom
above the cutoff,
thereby drastically reducing the undesirable (or
unphysical) sensitivity to short-distance physics.

\section{The \lowercase{$hep$} process}
In the notation of MSVKRB, the GT-amplitudes are given in
terms of the reduced matrix elements, $\overline{L}_1(q;A)$ and
$\overline{E}_1(q;A)$. Since these matrix
elements are related to each other
as $\overline{E}_1(q;A) \simeq \sqrt{2}\, \overline{L}_1(q;A)$,
with the exact equality holding at $q$=0, we consider here only
one of them, $\overline{L}_1(q;A)$. For the three exemplary values
of $\Lambda$, Table~\ref{TabL1A} gives the corresponding values of
$\overline{L}_1(q;A)$ at $q\equiv|\vq|$=19.2 MeV and zero c.m.
energy; for convenience, the values of $\dR$ in Eq.(\ref{dR}) are
also listed. We see from the table that the variation of the
two-body GT amplitude (row labelled ``2B-total'') is only
$\sim$10 \% for the range of $\Lambda$ under study.  Note that the
$\Lambda$-dependence in the total GT amplitude is
made more pronounced
by the drastic cancellation between the one-body
and two-body terms, but this amplified $\Lambda$-dependence still
lies within acceptable levels.

\begin{table}[htb]
\caption{\label{TabL1A}\protect Values of $\dR$ and
$\overline{L}_1(q;A)$ (in fm$^{3/2}$) for the $hep$ process
calculated as functions of the cutoff $\Lambda$. The individual
contributions from the one-body (1B) and two-body (2B) operators
are also listed.}
\begin{tabular}{c|rrr}
$\Lambda$ (MeV) & 500 & 600 & 800 \\ \hline
$\dR$      & $1.00 \pm 0.07$ & $1.78\pm 0.08$ & $3.90\pm 0.10$
\\ \hline
$\overline{L}_1(q;A)$ & $-0.032$ & $-0.029$ & $-0.022$ \\ \hline
1B                    & $-0.081$ & $-0.081$ & $-0.081$ \\ \hline
2B (without $\dR$)    & $0.093$  & $0.122$  & $0.166$  \\
2B ($\propto \dR$)    & $-0.044$ & $-0.070$ & $-0.107$ \\ \hline
2B-total              & $0.049$  & $0.052$  & $0.059$  \\
\end{tabular}
\end{table}

\begin{table}[hbt]
\caption{\label{TabS}\protect Contributions
to $S_{hep}(0)$ (in \Sunit)
from individual initial channels calculated
as functions of $\Lambda$.
The last column gives the results obtained
in MSVKRB.}
\begin{tabular}{c|ccc|c}
$\Lambda$ (MeV) & 500 & 600 & 800 & MSVKRB
\\ \hline
${}^1S_0$ & 0.02  &  0.02  &  0.02 & 0.02 \\
${}^3S_1$ & 7.00  &  6.37  &  4.30 & 6.38 \\
${}^3P_0$ & 0.67  &  0.66  &  0.66 & 0.82 \\
${}^1P_1$ & 0.85  &  0.88  &  0.91 & 1.00 \\
${}^3P_1$ & 0.34  &  0.34  &  0.34 & 0.30 \\
${}^3P_2$ & 1.06  &  1.06  &  1.06 & 0.97 \\ \hline
Total     & 9.95  &  9.37  &  7.32 & 9.64 \\
\end{tabular}
\end{table}

Table~\ref{TabS} shows the contribution to the $S$-factor,
at zero c.m.\  energy, from each initial channel. For comparison
we also give the results of MSVKRB for the AV18/UIX interaction.
It is noteworthy that for all the channels other than ${}^3S_1$,
the $\Lambda$-dependence is very small ($\lesssim 2$ \%). The
${}^3S_1$ channel is the most sensitive to short-distance physics
because the extraordinary suppression of the one-body GT
contribution makes more pronounced the chiral-non-protected nature
of the GT transition. In fact, the sensitivity of the ${}^3S_1$
channel to short-distance physics would be larger if  the
contribution of the $A^0$ term, which is rather sizable here
despite its generic $1/m$ suppression, were omitted. It is
therefore reassuring that the chiral-filter mechanism allows
reliable estimation of the $A^0$ term in this channel as well
(besides the $P$-wave channels), see~\cite{MSVKRB}.

Summarizing the results given in Table~\ref{TabS}, we
arrive at a prediction for the $hep$ $S$-factor\footnote{See
below for a possible caveat on the given
error estimate.}:
 \be
S_{hep}(0)=(8.6 \pm 1.3 ) \times \Sunit\,,\label{prediction} \ee
where the ``error" spans the range of the $\Lambda$-dependence for
$\Lambda$=500--800 MeV. This result should be compared to that
obtained by MSVKRB~\cite{MSVKRB}, $S_{hep}(0)=9.64 \times
\Sunit$.\footnote{ The earlier studies~\cite{CRSW91,SWPC92} were
based on less accurate variational wave functions than used here
and in MSVKRB; furthermore they did not include P-wave capture
contributions, which account for $\approx 40$ \% of the total
$S$-factor.}

The latest analysis of the Super-Kamiokande data
\cite{SK2001} gives an upper limit
of the solar $hep$ neutrino flux,
$\Phi(hep)^{\rm SK} < 40
\times 10^3$ cm$^{-2}$s$^{-1}$.
The standard solar model
\cite{BP2000} using the $hep$ $S$-factor
of MSVKRB~\cite{MSVKRB} predicts
$\Phi(hep)^{\rm SSM} = 9.4
\times 10^3$ cm$^{-2}$s$^{-1}$.
The use of the central value of our estimate,
Eq.(\ref{prediction}), of the $hep$ $S$-factor
would slightly lower $\Phi(hep)^{\rm SSM}$
but with the upper limit compatible with
$\Phi(hep)^{\rm SSM}$ in Ref.~\cite{BP2000}.
A significantly improved estimate of $S_{hep}(0)$
in Eq.(\ref{prediction})
is expected to be useful for further discussion
of the solar $hep$ problem.

To reduce the uncertainty in Eq.(\ref{prediction}), we need to
reduce the $\Lambda$-dependence in the two-body GT term. According
to a general {\it tenet} of EFT, the cutoff dependence should
diminish as higher order terms get included. In fact, the somewhat
rapid variation seen in $\dR$ (Table \ref{TabL1A}) and in the
${}^3S_1$ contribution to $S_{hep}(0)$ (Table \ref{TabS}) as
$\Lambda$ approaches 800 MeV may be an indication that there is
need for the explicit presence of the vector-mesons ($\rho$ and
$\omega$) with mass $m_V \lsim \Lambda$. This possible
insufficiency could be remedied to a certain extent by going to
higher orders. A preliminary study~\cite{PKMRhep} indicates that
it is indeed possible to reduce the $\Lambda$-dependence
significantly by including \nlo4 corrections. We expect that the
higher order correction would make the result for $\Lambda=800$
MeV closer to those for $\Lambda=500, 600$ MeV, bringing the MEEFT
results closer to what was obtained in MSVKRB. This possibility is
taken into account in the error estimate given in
Eq.(\ref{prediction}).

\section{Discussion}

It is worth emphasizing that the above MEEFT prediction
for $\delta_{\rm 2B}$ for the $pp$ process
is in line with the latest SNPA
results obtained in Ref.~\cite{TBDexp} (and mentioned earlier).
There too, the short range behavior of the axial MEC was
constrained by reproducing $\Gamma^t_\beta$.
The inherent model dependence of such a procedure
within the SNPA context was shown to be very weak simply because
at small inter-particle separations, where
MEC contributions are largest, the pair wave functions in
different nuclei are similar in shape and differ only by a scale
factor~\cite{forest96}.
As a consequence, the ratios of GT and $p$$p$-capture
matrix elements of different two-body current terms
are nearly the same,
and therefore a knowledge of their sum in the GT matrix element
is sufficient to predict their sum in the
$p$$p$-capture matrix element~\cite{TBDexp}.

In order to better understand how the present scheme works,
it is helpful to compare the $hep$ reaction with the
radiative $np$-capture.
The polarization observables in
$\vec{n}+\vec{p}\rightarrow d+\gamma$ are known
to be sensitive to the isoscalar M1 matrix element,
$M1S$,
and this amplitude has been extensively studied
in EFT \cite{npp,crs99}.
The similar features of the $hep$ GT amplitude
and $M1S$ are:
(i) the leading one-body contribution
is suppressed by the symmetries of the wave functions;
(ii) there is no soft-pion exchange contribution;
(iii) nonetheless,
short-range physics can be reliably subsumed
into a single contact term.
In the $\vec{n}\vec{p}$ case
the strength of this term can be determined
from the deuteron magnetic moment
(for a given value of $\Lambda$).
The calculation in Ref.~\cite{npp} demonstrates
that the $\Lambda$-dependence
in the contact term and that of the remaining terms
compensate each other so that
the total $M1S$ is stable against changes
in $\Lambda$.
This suggests that, if we go to higher orders,
the coefficient of the contact term
in question will be modified,
with part of its strength shifted to higher order terms;
however, the total physical amplitude will
remain essentially unchanged.
These features are quite similar to what we have found here
for the $hep$ GT amplitude.

Evaluating the matrix element of the leading-order one-body
operator in EFT with the use of realistic nuclear wave functions
is analogous to fixing parameters in an EFT Lagrangian (at a
given order) using empirical inputs~\cite{PKMR98};
the realistic wave functions in SNPA can be regarded as a
theoretical input that fits certain sets of observables. In
the present MEEFT scheme, we take the view that the same realistic
wave functions also provide a framework for reliably calculating
(finite-range) many-body corrections to the leading-order one-body
matrix element. The short-ranged part inherited from
the integrated out degrees of freedom
is regulated by the $\dR$ term.
This way of handling ``short-range correlation"
is analogous to Bogner {\it et al.}'s derivation \cite{kuo}
of ``$V_{low-k}$"
based on renormalization-group theory.
While our approach here is,
in certain cases, not in strict accordance with
the systematic power-counting scheme of EFT proper, nevertheless
the severity of this potential shortcoming should
vary from one case
to another (see discussion in Ref.~\cite{beaneetal2}).
For the $pp$ and $hep$ amplitudes under consideration,
the degree of $\Lambda$-dependence exhibited
by the numerical results does suggest
that deviations from rigorous power-counting cannot be too
significative. Indeed, this type of ``resilience" may also explain
why the SNPA calculation in Ref.~\cite{MSVKRB} gives a result very
similar to the present one. It is true that the two-body terms in
MSVKRB are not entirely in conformity with the chiral counting
scheme we are using here; some terms corresponding to chiral
orders higher than \nlo3 are included, while some other terms
which are \nlo3 in EFT are missing (see Appendix B.3).
Most importantly the
$\dR$-term -- that plays a crucial role here -- is omitted in
MKSVRB although heavy-meson exchange graphs may account for some
part of it. This formal problem, however, seems to be largely
overcome by the fact that also in MSVKRB a parameter (the axial
$N\Delta$ coupling strength) is adjusted to reproduce
$\Gamma_\beta^t$.

Not unrelated to the above issue of power-counting is the question
of consistency of embedding ``realistic" wave functions obtained
from ``realistic" potentials that are fitted $accurately$ to
experiments into an EFT framework with the currents obtained to a
given order of chiral perturbation theory. It is a well-known fact
that potentials that fit experiments are not necessarily unique.
The non-uniqueness resides however in the short-range part of the
potential, with the long-range part primarily governed by the pion
exchange. Let us suppose that one can calculate potentials to a
very high order in a consistent expansion (that is, consistent
with symmetries etc.). The structure of the potential would depend
on various aspects of the calculation. For instance, although they
all may fit equally well various experimental data such as e.g.,
nucleon-nucleon scattering, different regularizations would lead
to different potentials, the difference residing mainly in the
short-range part. 
One might worry that this non-uniqueness would
upset the basic premise of an EFT, rendering the predictions
untrustworthy.

Another intricate issue,
which is also connected to 
short-range physics,
is the off-shell ambiguity.
This problem should be absent 
in a formally consistent EFT. 
In MEEFT, however, 
we insert the current operators
derived from irreducible diagrams up to a
given chiral order 
between phenomenological (albeit realistic)
wave functions.
Since the inserted current
involves off-shell particles, there can in principle be 
terms other than those that have been included
in our approach. 
While those additional terms
that may be required to eliminate the off-shell dependence are
expected to be of higher order than \nlo3, 
this issue warrants a further examination.

To answer the above question with full rigor, 
much more work
is needed. 
However, partial and yet reasonably satsifactory answers
can be obtained from this work.
For chiral-filter protected processes, we have presented a clear
argument that the above-mentioned ambiguity does not matter at the
level of accuracy in question. The results listed in Table
\ref{TabS} for the $P$-wave capture (to which the chiral-protected
time component of the axial current contributes) demonstrate this
point. The question of short-distance ambiguity arises only for
chiral-unprotected processes like the GT transition. As already
explained, however, the $\dR$ regularization essentially removes
this ambiguity. The point is that the physics of the degrees of
freedom above the cutoff scale $\Lambda$ gets lodged in the
short-range $\dR$ term. In fixing this term as a function of
$\Lambda$ via the experimental value of $\Gamma_\beta^t$, one is
essentially incorporating the short-range correlations that render
low-energy physics insensitive to short-distance physics. 

As for the off-shell problem,
we note that for processes involving 
a long-wavelength external current --
such as the solar $pp$ and $hep$ reactions --
the off-shell ambiguity should be small,
so long as one uses high-quality phenomenological
wave functions that accurately describe
processes without the external current.
The wave functions used
here describe with high accuracy a rich ensemble of data available
for the systems in question; they describe very well the
three-nucleon scattering states, and furthermore, the $n^3$He
elastic scattering cross section as well as the coherent
scattering length calculated with these wave functions are in
excellent agreement with the experiments.
What is involved here
seems to be a generic feature. 
A similar stabilizing mechanism is at
work when Bogner {\it et al}.~\cite{kuo} arrive at a unique
effective force $V_{low-k}$ by integrating out the
high-energy/momentum components contained in various ``realistic"
potentials. 
Nuclear physics calculations done with this effective
force~\cite{kuo2} have much in common with the MEEFT calculation
described here.
Furthermore, we remark that different off-shell
properties reflect different choices of 
the field variables
and that, for each choice, the LECs need to be
readjusted.
It is in principle possible
to choose the field variables in such a manner  
that off-shell contributions become highly suppressed.
We are essentially adopting this particular choice 
by using the forms of the transition operators
described above and adjusting the corresponding LEC, $\dR$,
to reproduce $\Gamma^t_\beta$.

A possible approach that is formally consistent with systematic
power counting is the pionless EFT based on the power divergence
subtraction (PDS) scheme \cite{pds}
(for a recent review, see Ref.~\cite{seattle}), which has been applied
to the $pp$ fusion \cite{bc01}. Due to the fact that this scheme
also involves one unknown low-energy constant, PDS has not so far
led to a definite prediction on the $pp$ fusion rate. The problem
is that this approach cannot be readily extended to systems with
$A\ge3$, in particular to electroweak transition amplitudes in
these systems. What is lacking presently is a method to
correlate in a unified framework the observables
in different nuclei (different mass numbers).
This limitation
keeps one from exploiting the experimental data available for the
$A\ge3$ nuclei to fix unknown LEC.  Apart from the basic
problem of organizing chiral expansion for complex nuclei from
``first-principles", a plethora of parameters involved would
present a major obstacle. (For recent efforts in this approach,
see Ref.~\cite{seattle,bedaque02} and references given therein.)
This difficulty is expected to be particularly pronounced for the
$hep$ reaction.

There has been a series of intensive studies by the J\"ulich Group
to extend EFT calculations in the Weinberg scheme to systems with
three or more nucleons~\cite{epeetal00}. The relationship between
this approach and the phenomenological potential approach has been
examined in great detail. This line of study, however, has been so
far limited to nuclear observables that do {\it not} involve the
electroweak currents. An extension of the formalism developed in
Ref.~\cite{epeetal00} to electroweak transitions should be
extremely useful.

\section*{Acknowledgement}
TSP and KK thank S. Ando and F. Myhrer for discussions. MR is
grateful for clarifying discussions with Gerry Brown on the notion
of ``double decimation" in nuclear structure theory. The work of
TSP and KK is supported in part by the U.S. National Science
Foundation, Grant No.~PHY-9900756 and No.~INT-9730847. The work of
RS is supported by DOE contract No.~DE-AC05-84ER40150, under which
the Southeastern Universities Research Association (SURA) operates
the Thomas Jefferson National Accelerator Facility. The work of
DPM is supported in part by KOSEF Grant 1999-2-111-005-5 and KSF
Grant 2000-015-DP0072. MR acknowledges the hospitality of the
Physics Departments of the Seoul National University and Yonsei
University, where his work was partially supported by Brain Korea
21 in 2001. TSP and KK would like to thank Korea Institute for
Advanced Study for the hospitality while part of this paper was
being written. Some of the calculations were made possible by
grants of computing time from the National Energy Research
Supercomputer Center in Livermore.

\appendix



\section{Gamow-Teller Operators}

\setcounter{equation}{0}
\renewcommand{\theequation}{\mbox{A.\arabic{equation}}}
The aim of this and subsequent Appendices
is to provide some technical details that have been
left out in the main text. 
The readers who are not interested in the details 
of our calculation can safely skip these Appendices
without missing the essential points
of our results.

We decompose the axial current into $n_B$-body operators as
 \be
A^{\mu,a} &=& A^{\mu,a}_{\rm 1B} + A^{\mu,a}_{\rm 2B} +
A^{\mu,a}_{\rm 3B} + \cdots
\nonumber \\
&=& \sum_l A_l^{\mu,a} + \sum_{l<m} A_{lm}^{\mu,a} + \sum_{l<m<n}
A_{lmn}^{\mu,a}\nonumber\\
&& + \cdots,
 \ee where $(l,\, m,\, n)$ are particle
indices. The one-body operator can be read from
 \be &&\langle N(p')
| A^{\mu,a}(x=0) | N(p) \rangle \nonumber\\
&&=  - {\bar u}(p') \left[
  G_{\!A}(k^2) \gamma^\mu \gamma_5
  - \frac{G_{\!P\!}(k^2)}{2 m_N} k^\mu \gamma_5
  \right] \frac{\tau^a}{2} u(p),
\label{A1B} \ee where $u(p)$ is a four-component Dirac spinor of
momentum $p$, and $k^\mu$=$(p - p')^\mu$ is the momentum carried
by the lepton pair. $G_{\!A\!}(k^2)$ and $G_{\!P\!}(k^2)$ are the
axial and induced pseudoscalar form factors, respectively. Note
that $k^\mu$=${\cal O}(Q^2/m_N)$, while $\vp$=${\cal O}(Q)$ and
$\vp\,'$=${\cal O}(Q)$. Thus the contribution from $G_P$ term is
of ${\cal O}(Q^4)$. In getting the non-relativistic operators from
the above relativistic form factors, we should also take into
account the wave function normalization. 
The resulting one-body
operator up to ${\cal O}(Q^3)$ then reads \be \vA_l^a &=&
 - \frac{\tau_l^a}{2}\, g_A\, \e^{-i \vr_l\cdot \vq}\,\left[
  \vs_l + \frac{\vbp_l\,\vs_l\cdot \vbp_l - \vs_l\,\vbp_l^2}{2
  m_N^2}\right.\nonumber\\
 &&\left. + \frac{i \vq\times\vp_l}{4 m_N^2}
 + {\cal O}(\frac{Q^4}{m_N^4}) \right].
\label{A1Bnon}\ee 
This expression has appeared in eq.(\ref{J1Bnon}).

In the following subsections, 
we derive all the two-body GT operators up to \nlo4
and leading three-body GT operators .

\subsection{Two-body GT}
\begin{figure}[htbp]
\centerline{\epsfig{file=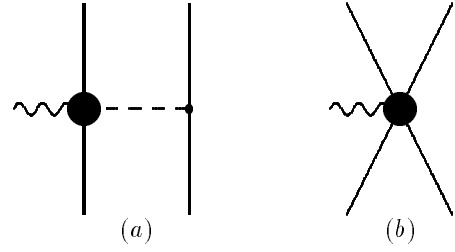,width=6cm}} \vskip 0.5cm
\caption{\protect A one-pion pole diagram, ($a$),
responsible for $\vA_{lm}^a(1\pi)$, and a short-range contribution
diagram, ($b$), responsible for $\vA_{lm}^a(2\pi)$. The solid
circles include counter-term insertions and (one-particle
irreducible) loop corrections. The wiggly line stands for the
external field (current) and the dashed line for the pion.
One-loop corrections of the relevant orders for the pion
propagator and the $\pi NN$ vertex need to be included.}
\label{2Bfig}
\end{figure}

\indent There are no two-body GT diagrams that involve only
$\nu_i=0$ vertices, because the ${\cal A}^{i,a}\pi NN$ vertex is
kinematically suppressed, and there is no four-fermion contact
contribution at LO ($\nu_i=0$). As a consequence, the two-body GT
operator starts at $\nu=3$.
The two-body GT operator at threshold ($q^\mu\rightarrow 0$) was
given up to \nlo3 in Ref.~\cite{PJM}. 
We extend here that analysis to
include all contributions up to \nlo4. To this end, it is useful
to decompose $\vA_{lm}^a$ as \be \vA_{lm}^a= \vA_{lm}^a(1\pi) +
\vA_{lm}^a(2\pi), \label{decompose} \ee where $\vA_{lm}^a(1\pi)$
represents the contributions of the one-pion pole part and
$\vA_{lm}^a(2\pi)$ stands for the remaining short-ranged part.
Generic diagrams for $\vA_{lm}^a(1\pi)$ and $\vA_{lm}^a(2\pi)$ are
shown in Fig.~\ref{2Bfig}.

We now list all the two-body GT operators belonging to $\nu$=3 and
$\nu$=4.

\vspace*{0.2cm}
\noindent $\bullet$ $\nu=3$:

This contribution comes from tree graphs (one-pion-exchange and
contact) with a $\nu_i=1$ vertex. The resulting GT operators,
given in Ref.~\cite{PJM}, are of the form
 \be \vA_{12}^{a:\nu3}(1\pi)
 &=& \frac{g_A}{2 m_N f_\pi^2} \left[
 \frac{i}{2} (\vtau_1\times \vtau_2)^a \vbp_1
 + 2 \hat c_3 \tau_2^a \vk_2\right.\nonumber\\
 &+&\left.
  (\hat c_4 + \frac14) (\vtau_1\times \vtau_2)^a \vs_1\times \vk_2
 \right.
\nonumber \\
&+&\left.
 \frac{1+c_6}{4}
  (\vtau_1\times \vtau_2)^a \vs_1\times \vq
 \right] \frac{\vs_2\cdot\vk_2}{\vk_2^2 + m_\pi^2}\nonumber\\
 &+& (1\leftrightarrow 2),\label{23}
\\
\vA_{12}^{a:\nu3}(2\pi)
 &=& \frac{g_A}{m_N f_\pi^2}
 \left[ \hat d_1 (\tau_1^a \vs_1 + \tau_2^a
 \vs_2)\right.\nonumber\\
 &+&\left. \hat d_2 (\vtau_1\times \vtau_2)^a \vs_1\times \vs_2
 \right]\,.
\label{vAnu3}\ee Although there are two unknown parameters,
$\hat d_1$ and $\hat d_2$,
it turns out that the Fermi-Dirac statistics
effectively reduces the number of unknowns to one. We will come
back to this important point later.

\vskip 0.5cm \noindent $\bullet$ $\nu=4$:

Tree graphs with $\sum_i \nu_i=2$ and one-loop graphs with $\sum_i
\nu_i=0$ enter at this order. Since there is no $\pi NN$ vertex
with $\nu_i=1$, a $\nu=4$ tree graph should have either $\vA
(NN)^2$ or $\vA \pi NN$ vertex with $\nu_i$=2. We can, however,
exclude either possibility. The absence of $\vA \pi NN$ vertex at
$\nu_i$=2 can be ascertained by consulting a complete list of
terms that appear in the
\nlo2 Lagrangian given in Ref.~\cite{fettes}. As for the $\nu_i=2$ $\vA
(NN)^2$ vertex for the two-nucleon sector, a complete list is not
available yet. We therefore resort to parity selection rules. Our
vertex must have one $\Delta_\mu$ and one $D_\mu$ involving four
nucleon fields. These operators should not be contracted with the
four-velocity $v^\mu$, for otherwise the actual chiral index would
acquire an extra power of $Q$. We can easily show, however, that
it is impossible to construct a parity-even Lorentz scalar with
one $\Delta_\mu$, one $D_\mu$ and arbitrary numbers of $S^\mu$ and
$\epsilon^{\mu\nu\alpha\beta}$. Introducing an operator of the
$\del_\mu A_\nu - \del_\nu A_\mu$ type instead of $\Delta_\mu$ and
$D_\mu$ does not help either.
These observations lead us to conclude that no
divergences occur in the relevant loops and,
more importantly, that
no new parameters appear at $\nu$=4.

\begin{figure}[htbp]
\centerline{\epsfig{file=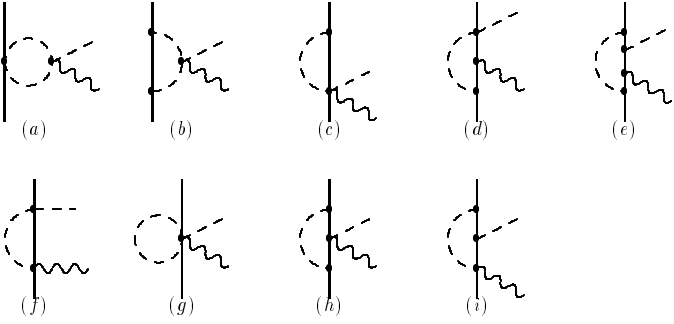,width=9cm}}
 \vskip 0.5cm
\caption[Api]{\protect \small One loop diagrams that contribute to
the ${\cal A}^\mu \pi NN$ vertex. Only the first five diagrams
$(a)-(e)$ contribute to $\vA$.} \label{Apifig}
\end{figure}

The one-pion-exchange contribution can be read off from 1-loop
corrections to the $\Gamma_{{\cal A}\pi}^{\mu,ab}$ vertex; the
relevant diagrams are shown in Fig.~\ref{Apifig}. We note that
only the first five diagrams ($a$)-($e$) contribute to $\vA$.
Using the expressions given in Ref.~\cite{axialch}, where all the
one-loop diagrams have been calculated, we find
 \be &&\vA_{12}^{a:\nu4}(1\pi)
= - \frac{g_A^3}{32\pi f_\pi^4} \nonumber\\
 &&\otimes\left[
 (\vtau_1\!\times\!\vtau_2)^a\!
 \left( \vs_1\!\times\!(\vk_2\!+\!\vq) D_1(k_1)
  \!+\!\frac{g_A^2}{8} \vs_1\!\times\!\vk_2\,m_\pi\!\right)
\right.
\nonumber \\
&& -\ \tau_2^a \left(
    (\vq \!+\! 3 \vk_2)
      \left[ D_1(k_1)
      \!+\!\frac{1}{3} \vk_1^2 D_2(k_1) \right]
 \! + \!\frac{9 g_A^2}{8} \vk_2\,m_\pi \!\right)
\nonumber \\
&& \left. \frac{}{} \right] \frac{\vs_2\cdot \vk_2}{m_\pi^2 +
\vk_2^2}
 + (1 \leftrightarrow 2),
\label{vAnu41pi}\ee
where $k_l=|\vk_l|$ ($l=1,\,2$), and
$D_i(k)$'s are defined as \be
 D_1(k) &=& \int_0^1\!dz\, M_{zk},
\nonumber \\
D_2(k) &=& \int_0^1\!dz\, \frac{z\bar z}{M_{zk}},
\nonumber \\
D_3(k) &=& \int_0^1\!dz\, \left(
 - \frac{z\bar z\, \vk^2}{M_{zk}} - 5 M_{zk} \right)
\nonumber \\
&=& 4 \int_0^1\!dz\, \left( -\frac32 M_{zk} + \frac{m_\pi^2}{4
M_{zk}}\right),
\nonumber \\
D_4(k) &=&
 \int_0^1\!dz\, \left(
 - \frac{(z \bar z)^2 \vk^2}{M_{zk}^3}
 + 7 \frac{z \bar z}{M_{zk}}
 - \frac{1}{M_{zk}}\right)
\nonumber \\
&=& 4\int_0^1\!dz\, \left(
 \frac{z \bar z}{2 M_{zk}}
 + \frac{z \bar z m_\pi^2}{4 M_{zk}^3}
 - (\frac14 - z \bar z ) \frac{1}{M_{zk}}\right),
\nonumber \\
D_5(k) &=& \int_0^1\!dz\,
 \frac{1}{M_{zk}},
\nonumber \\
D_6(k) &=& \int_0^1\!dz\,
 (\frac14 - z \bar z ) \frac{1}{M_{zk}}.
\label{Di}\ee Here $k\equiv |\vk|$,
$\bar z\equiv 1-z$, and $M_{zk}\equiv
\sqrt{m_\pi^2 + z \bar z\,{\vk}^2}$.

The integrations over $z$ can be done analytically, resulting in
\be D_1(k) &=& \frac{m_\pi}{2} + \frac{4 m_\pi^2+k^2}{4 k}
\Theta_k,
\nonumber \\
D_2(k) &=& \frac{m_\pi}{2 k^2} - \frac{4 m_\pi^2-k^2}{4 k^3}
\Theta_k,
\nonumber \\
D_3(k) &=& -3 m_\pi - \frac{8 m_\pi^2 + 3 k^2}{2 k}  \Theta_k,
\nonumber \\
D_4(k) &=& \frac{1}{2 k^3} \left[
 \frac{2 m_\pi k (8 m_\pi^2+ 3 k^2)}{4 m_\pi^2+k^2}
 - (8 m_\pi^2+ k^2) \Theta_k\right],
\nonumber \\
D_5(k) &=&
 \frac{2}{k} \Theta_k,
\nonumber \\
D_6(k) &=& - \frac{m_\pi k}{2 k^3} +
 \frac{4 m_\pi^2+ k^2}{4 k^3} \Theta_k
\label{DiA} \ee where $k\equiv |\vk|$ and \be \Theta_k \equiv
\tan^{-1}\frac{k}{2 m_\pi} \ee with $-\frac{\pi}{2} < \Theta_k <
\frac{\pi}{2}$.

Note that the one-pion-pole contributions can be absorbed into
$\vA_{lm}^{a:\nu3}(1\pi)$ (given in Eq.(\ref{23})) by renormalizing
$\hat c_3$ and $\hat c_4$, \be \hat c_3 &\rightarrow& \hat c_3^R
 \equiv \hat c_3 +
   \frac{m_N g_A^2}{32\pi f_\pi^2} \left[
     \tilde D_{1\pi}+\frac{9 g_A^2}{8} m_\pi \right]\nonumber\\
 &\simeq& \hat c_3 + 1.0334,
\nonumber \\
\hat c_4 &\rightarrow& \hat c_4^R
 \equiv \hat c_4 -
   \frac{m_N g_A^2}{32\pi f_\pi^2} \left[
     2 D_{1\pi}+\frac{g_A^2}{8} m_\pi \right]\nonumber\\
 &\simeq& \hat c_4 - 0.4821 \,.
\label{c3c4R}\ee where \be D_{1\pi}&\equiv& \left.
D_1(k)\right|_{k^2=-m_\pi^2} = \frac{m_\pi}{4} \left[2 + 3
\tanh^{-1}\frac12\right],
\nonumber \\
\tilde D_{1\pi}&\equiv& \left.
 3 D_1(k) + k^2 D_2(k) \right|_{k^2=-m_\pi^2}\nonumber\\
&=& m_\pi \left[2 + \tanh^{-1}\frac12\right]. \ee

\begin{figure}[htbp]
\centerline{\epsfig{file=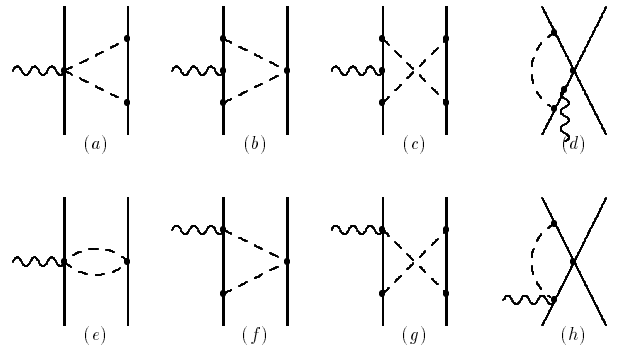,width=8cm}}
 \vskip 0.5cm
\caption[Api]{\protect \small One loop diagrams for the
$\vA_{lm}^{a:\nu4}(2\pi)$: the first four diagrams $(a)-(d)$
contribute to the space part of the axial-vector (GT) and the
remaining diagrams $(f)-(h)$ to the axial-charge current. The graph
$(e)$ vanishes.} \label{2pifig}
\end{figure}

For the two-pion contribution $\vA_{lm}^{a:\nu4}(2\pi)$, the
relevant one-loop graphs are shown in Fig.~\ref{2pifig}. Among the
diagrams in the figure, only the first four graphs, $(a)-(d)$, can
contribute to the GT operator; $(e)$ is identically zero due to
isospin symmetry, and the remaining graphs, $(f)-(h)$, contribute
only to $A^0$.

\vspace*{1cm}
As mentioned, the four diagrams $(a)-(d)$ are all
ultraviolet-finite. The first three graphs give
 \be
\vA_{12}^a(2\pi:a) &=&
 - \frac{g_A^3}{64\pi f_\pi^4} \vs_1 \tau_1^a
  \left[3 D_1(k_2) + \vk_2^2 D_2(k_2)\right]\nonumber\\
 &+& (1\leftrightarrow 2),
\nonumber \\
\vA_{12}^a(2\pi:b) &=&
 - \frac{g_A^3}{128\pi f_\pi^4} \tau_2^a
  \left[\vs_1 D_1(k_2)\right.\nonumber\\
  &-&\left. (2 \vk_2\, \vs_1\cdot\vk_2
      - \vs_1\,\vk_2^2) D_2(k_2)\right]\nonumber\\
 &+& (1\leftrightarrow 2),
\nonumber \\
\vA_{12}^a(2\pi:c) &=&
 - \frac{g_A^5}{1024\pi f_\pi^4} (\tau_1^a + 2\tau_2^a)
  \left[\vs_1 D_3(k_2)\right.\nonumber\\
     &+&\left. (2 \vk_2\, \vs_1\cdot\vk_2 - \vs_1\,\vk_2^2) D_4(k_2)
\right. \nonumber \\ &+& \left.
     (\vs_2\, \vk_2^2 - \vk_2\, \vs_2\cdot\vk_2)
     D_5(k_2)\right]\nonumber\\
 &+& (1\leftrightarrow 2)
\ee while the fourth diagram $(d)$ gives
 \be \vA_{12}^a(2\pi:d)=
 &-&\frac{g_A^3 m_\pi}{64\pi f_\pi^2}
\sum_A
 \tau_1^b \sigma_1^j
 \left\{C_A \Gamma_A \Gamma_A,\,\tau_1^a \vs_1\right\}
 \tau_1^b \sigma_1^j \nonumber\\
 &+& (1\leftrightarrow 2) \ .
\label{A3d1}\ee The summation here is taken over all possible
combinations of spin-isospin operators (with no derivatives) that
figure in the nucleon-nucleon interactions. Using a generic
expression
\be \sum_A C_A \Gamma_A \Gamma_A
 &=& X_1 + \vs_1\cdot \vs_2\, X_\sigma
 + \vtau_1\cdot \vtau_2\, X_\tau\nonumber\\
 &+& \vs_1\cdot \vs_2\, \vtau_1\cdot\vtau_2\,X_{\sigma\tau}
\label{CGG}\ee where $X_1,\,X_\sigma,\,X_\tau,\,X_{\sigma\tau}$
are constants that characterize the LO short-range nuclear forces,
we can write
 \be \vA_{12}^a(2\pi:d) &=&
 - \frac{g_A^3 m_\pi}{32\pi f_\pi^2} \left[
  (-3 X_1 + 9 X_{\sigma\tau}) (\tau_1^a \vs_1 + \tau_2^a
  \vs_2)\right.\nonumber\\
 &&\left. \ \ \ +\ (-2 X_{\sigma\tau}) (\vtau_1\times\vtau_2)^a (\vs_1\times\vs_2)
\right. \nonumber \\
&&\left. \ \ \
 +\ (9 X_\sigma - 3 X_\tau) (\tau_1^a \vs_2 + \tau_2^a \vs_1)
  \right]\,.
\label{A3d}\ee We demonstrate below that $X$'s can all be absorbed
into the parameters, $\hat d$'s.

\subsection{Three-body GT}
\begin{figure}[htbp]
\centerline{\epsfig{file=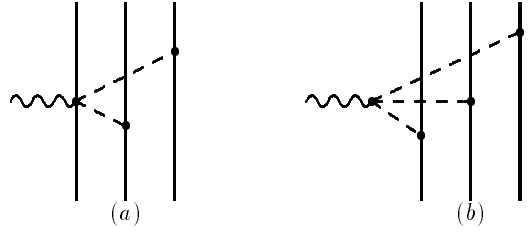,width=7cm}}
 \vskip 0.5cm
\caption{\protect Diagrams for the three-body
GT operator. All other diagrams are higher order than \nlo4 except
for the crossed diagrams of ($a$) -- crossed with respect to the
particle indices.} \label{3Bfig}
\end{figure}

\indent The three-body GT operators up to \nlo4 come from the two
diagrams given in Fig.~\ref{3Bfig}. They contain only $\nu_i=0$
vertices, and their contributions read \be \vA_{123}^a &=&
- \sum_{{\rm cycle}(123)}\frac{g_A^3}{16 f_\pi^4}
\nonumber \\
&&\otimes\
 (2 \tau_1^a\, \vtau_2\cdot\vtau_3
 -  \tau_2^a\, \vtau_3\cdot\vtau_1
 - \tau_3^a\, \vtau_1\cdot\vtau_2)
\nonumber \\
&&\otimes\
 \left( \vs_1 - \frac{4}{3} \frac{\vk_1\, \vs_1\cdot \vk_1}{\vk_1^2+m_\pi^2}
 \right)
 \frac{\vs_2\cdot \vk_2}{\vk_2^2+m_\pi^2}
 \frac{\vs_3\cdot \vk_3}{\vk_3^2+m_\pi^2}\,,
\label{AIII}\ee where \be \sum_{{\rm cycle}(lmn)} f_{lmn}\equiv
 f_{lmn} + f_{mnl} + f_{nlm}.
\ee

\vspace*{1cm}

\subsection{Comparison with SNPA exchange currents }
\indent The meson-exchange currents in SNPA~\cite{cr71,towner} are
based on one-boson exchange diagrams involving those bosons which
are responsible for the phenomenological nuclear forces in the
context of one-boson-exchange models. This framework does not have
direct contact with chiral counting.
We give here a detailed comparison between the transition
operators used in SNPA and those dictated by $\chi$PT. Among the most
elaborate SNPA operators are the ones used in
CRSW91~\cite{CRSW91}; these operators were derived by
Towner~\cite{towner} based on a phenomenological
Lagrangian~\cite{IT79} which satisfies CVC, PCAC and current
algebra. We consider the  SNPA operators used in CRSW91 as a
representative. It will turn out that there are substantial
differences between the SNPA and $\chi$PT operators in both the
long-range and short-range parts.

In CRSW91 the heavy particles, $\rho$ and $\Delta$, are treated as
explicit degrees of freedom.\footnote{ The one-body operators are
not sensitive to these additional ingredients as long as the
single-nucleon parameters are determined empirically.  We note,
however, that CRSW91 includes only the leading-order GT operator,
$\frac{\tau^a}{2} g_A \vs$, without taking into account
relativistic corrections.} To examine the roles of these heavy
particles in the context of the present comparison, we divide the
two-body currents in CRSW91 into two families:
 \be
\vA^a&=& \vA^a_I + \vA^a_{II}\\
&\equiv& \left[ \vA^a(\Delta\pi) +
 \vA^a(\pi\rho) + \vA^a(\pi S)\right]\nonumber\\
 &&\ + \left[\vA^a(\Delta\rho) + \vA^a(\rho S)\right],
\ee where the ``S" stands for ``seagull". $\vA^a_I$ and
$\vA^a_{II}$ can be associated, respectively, with
$\vA_{lm}^a(1\pi)$ and $\vA_{lm}^a(2\pi)$ in Eq.(\ref{decompose}).
The expression for $\vA^a_I$ is\footnote{ The pseudovector $\pi
NN$ coupling constant $f_{\pi\!N\!N}$ used in CRSW91 is related to
the quantities used here as
 \be \frac{f_{\pi\! NN}}{m_\pi} =
\frac{g_{\pi N\!N}}{2 m_N}
 = \frac{g_A}{2 f_\pi}.
\ee For the last equality, the Goldberger-Treiman relation has
been used. } \be
&& \vA^a_I = \frac{g_A}{2 m_N f_\pi^2} \left\{ \frac{}{} \right.
\nonumber \\
&& -\ \frac{4}{25} g_A^2 I_1 \frac{m_N}{m_\Delta\! - \!m_N}
  \calR_\pi^2(\vk_2)
  \left[ 4 \tau_2^a\vk_2 - (\vtau_1\times\vtau_2)^a\,
  \vs_1\!\times\!\vk_2
  \right]
\nonumber \\
&& - \frac{I_2}{4} \calR_\rho(\vk_1)
 \calR_\pi(\vk_2) \frac{m_\rho^2}{m_\rho^2 + \vk_1^2}
  (\vtau_1\!\times\!\vtau_2)^a\left[
  (1\!+\!\kappa) \,\vs_1\!\times\!\vk_1\!-\!2i \vbp_1\right]
\nonumber \\
&& + \ \frac{I_1}{4} g_A^2 \calR_\pi^2(\vk_2) \left[
  (\vtau_1\!\times\!\vtau_2)^a\, \vs_1\!\times\!\vk_2
 - \tau_2^a ( -\vq \!+\!2i \vs_1\!\times\!\vbp_1) \right]
\nonumber \\
&& \left. \frac{}{} \right\}
\frac{\vs_2\cdot\vk_2}{m_\pi^2 +\vk_2^2}
 + ( 1 \leftrightarrow 2)
\label{AijOld} \ee with \be {\cal R}_\pi(\vk) \equiv
 \frac{\Lambda_\pi^2 - m_\pi^2}{\Lambda_\pi^2 + \vk^2}\,,
\;\;\;\;\;\;{\cal R}_\rho(\vk) \equiv
 \frac{\Lambda_\rho^2 - m_\rho^2}{\Lambda_\rho^2 + \vk^2},
\ee where $m_\Delta\simeq 1232\ \MeV$, $\kappa\simeq 6.6$, and
$g_\rho \simeq 2.50$ is the $\rho NN$ coupling constant;
$\Lambda_{\pi}$ ($\Lambda_\rho$) is a cutoff parameter
characterizing the $\pi NN$ ($\rho NN$) coupling form factor. We
have defined $I_1$ and $I_2$ by \be I_1 \equiv \frac{4 f_\pi^2
f_{\pi NN}^2}{g_A^2 m_\pi^2} = \frac{f_{\pi NN}^2}{m_\pi^2}
\!\cdot\!
    \left(\frac{g_A^2}{4 f_\pi^2}\right)^{\!\!-1}\!\!,
\ \ I_2 \equiv \frac{8 g_\rho^2 f_\pi^2}{m_\rho^2}. \ee We note
that $I_1=1$ if we assume the Goldberger-Treiman relation, and
$I_2=1$ if the KSRF relation holds.\footnote{ Since $g_\rho$ used
in CRSW91 is half of the conventional one, the KSRF relation here
reads $2 (2 g_\rho)^2\!f_\pi^2 = m_\rho^2$, rather than $2 g_\rho^2
f_\pi^2 = m_\rho^2$.} The above equation should be compared with
$\vA_{lm}^{a:\nu3}(1\pi)$ in Eq.(\ref{vAnu3}). A little exercise shows
that, while the currents $\vA(\pi\Delta)$ and $\vA(\pi\rho)$ can
be related to certain currents in $\chi$PT, $\vA(\pi S)$ has no $\chi$PT
counterpart to the order considered here. A possible explanation
for the occurrence of this ``extra" term in SNPA is that $\vA(\pi
S)$ arises as a ``recoil" term associated with the use of the
pseudoscalar coupling. A $\chi$PT analog of $\vA(\pi S)$ would be a
$1/m_N$ term, but this term should be substantially suppressed;
hence a term like $\vA(\pi S)$ should be absent in chirally
invariant theory. Comparison of the coefficients of
$(\vtau_1\times \vtau_2)^a \vbp_1$, $\tau_2^a \vk_2$,
$(\vtau_1\times \vtau_2)^a \vs_1\times \vk_2$ and $(\vtau_1\times
\vtau_2)^a \vs_1\times \vq$ leads to the following correspondence
between $\chi$PT (left-hand side) and CRSW91 (right-hand side):
\be 1
&\leftrightarrow& I_2 \calR_\rho(\vk_1) \calR_\pi(\vk_2)
\frac{m_\rho^2}{m_\rho^2 + \vk_1^2},
\\
\hat c_3 &\leftrightarrow& - \frac{8}{25} g_A^2 I_1
\frac{m_N}{m_\Delta - m_N}
  \calR_\pi^2(\vk_j),
\\
\hat c_4 +\frac14 &\leftrightarrow& \frac{4}{25} g_A^2 I_1
\frac{m_N}{m_\Delta - m_N}
  \calR_\pi^2(\vk_j)\nonumber\\
&&+ I_2 \calR_\rho(\vk_1) \calR_\pi(\vk_2)
\frac{m_\rho^2}{m_\rho^2 + \vk_1^2} \frac{1+\kappa}{4},
\\
1 + c_6 &\leftrightarrow&
 I_2 \calR_\rho(\vk_1) \calR_\pi(\vk_2)
\frac{m_\rho^2}{m_\rho^2 + \vk_1^2} \,(1+\kappa).
 \ee
The presence of the momentum-dependence in ${\cal R}$'s and the
$\rho$-meson propagator prevents us from going beyond this
correspondence. To proceed, however, we may consider the
approximation \be 1\approx I_2 \calR_\rho(\vk_1) \calR_\pi(\vk_2)
\frac{m_\rho^2} {m_\rho^2 + \vk_1^2} \approx I_1
\calR_\pi^2(\vk_j). \ee We then find \be {\hat c}_3^{\rm CRSW} &=&
- \frac{8}{25} g_A^2 \frac{m_N}{m_\Delta - m_N}
 \simeq - 1.633,
\\
{\hat c}_4^{\rm CRSW} &=& - \frac12 {\hat c}_3^{\rm CRSW} +
\frac{\kappa}{4} \simeq 2.467,
\\
c_6^{\rm CRSW} &=& \kappa \simeq 6.6\, . \ee It is informative to
compare these values of ${\hat c}$'s with those obtained in a
resonance-exchange saturation analysis by Bernard, Kaiser and
Meissner (BKM)~\cite{bernard}. We note that the two approaches
give very different results for the $\Delta$-resonance
contributions. CRSW91 used the quark model value for the ratio,
$g_{AN\Delta}/g_{\pi N \Delta}$, the accuracy of which is rather
difficult to assess. Meanwhile, the resonance-saturation
calculation suffers from ambiguity related to the so-called
off-shell parameter $Z$. Considering these uncertainties, it is
perhaps not too surprising that BKM's estimate of the $\Delta$
contribution to ${\hat c}_3$, $|{\hat c}_3^\Delta|$=3.59, is $2.2$
times larger than the  estimate in CRSW91.
We also note that, while CRSW91
only includes the $\Delta$ and $\rho$-meson contributions, BKM's
calculation contains the contributions from scalar-meson and Roper
exchanges as well. According to BMK, \be {\hat c}_3^\Delta + {\hat
c}_3^{\rm scalar} + {\hat c}_3^{\rm Roper} &=& -3.59 -1.31 -0.06 =
-4.96,
\nonumber \\
{\hat c}_4^\Delta + {\hat c}_4^\rho + {\hat c}_4^{\rm Roper} &=&
1.80 + 1.53 + 0.11 = 3.44. \ee Thus the contributions of the
scalar-meson exchange are substantial. What is significant for our
calculation is the fact that the coefficients $c$'s can be
extracted directly from the $\pi N$ scattering data
\cite{bernard,fettes}. The most recent analysis~\cite{fettes}
gives \be {\hat c}_3 &=& (-5.58 \pm 0.08,\ -5.49\pm 0.01,\
-5.82\pm 0.08),
\nonumber \\
{\hat c}_4 &=& (3.26\pm 0.05,\ 3.29\pm 0.01,\ 3.30 \pm 0.04).
\label{c34} \ee These values are in reasonable agreement with
those obtained in the resonance saturation approach.
We should note, however, that the results in Eq.(\ref{c34}) belong
to an \nlo4 calculation wherein $\Delta(1232)$ as well as other
massive degrees of freedom have been integrated out. The explicit
inclusion of $\Delta(1232)$ would modify the values of $\hat c$'s,
because the $\Delta$ contribution to the $\hat c$'s should now be
excluded. We also should pay attention to a similar modification
of the LECs as we move from \nlo4 to \nlo3.
Although the difference between $\hat c$'s obtained in an \nlo3
calculation and those obtained in \nlo4 are of order of $Q^4$ and
hence can in principle be neglected in an \nlo3 calculation,
it is more natural and safer to use in our present calculation the
values obtained in an \nlo3 analysis \cite{bernard}, \be \nlo3:\ \
\ \hat c_3 &=& -4.96 \pm 0.23,
\nonumber \\
\hat c_4 &=& 3.40 \pm 0.09\,. \label{c3c4NLO}\ee The precise value
of $c_6$ is unimportant in the present context, since it is
suppressed by the kinematic factor $|\vq|$. In any event, the
results of BMK and CRSW91 are close
to each other, $c_6\simeq 5.83$.

We now discuss the short-ranged currents, $\vA^a_{II}=
\vA(\rho\Delta) + \vA(\rho S)$. According to CRSW91, the dominant
term in $\vA^a_{II}$ is of the form
 \be &&\vA^a(\rho\Delta) =
 \frac{g_A}{2m_N f_\pi^2}\,
 I_2 \frac{(1+\kappa)^2}{50 m_N (m_\Delta - m_N)}
 \calR^2_\rho(\vk_2) \frac{m_\rho^2}{m_\rho^2 + \vk_2^2}
\nonumber \\
&&
 \otimes\left[ 4 \tau_2^a (\vs_2\times\vk_2)\times\vk_2
 - (\vtau_1\times\vtau_2)^a \,\vs_1\times
  \left[(\vs_2\times\vk_2)\times\vk_2\right] \right]\nonumber\\
 &&+ (1 \leftrightarrow 2).
\ee It should be noted, however, that this term belongs to \nlo{5}
in chiral counting, and therefore its inclusion in CRSW91
constitutes a deviation from $\chi$PT. Although a particular
\nlo{5} term may give an appreciable contribution (see below),
there are many terms of the same order, including multi-loop
diagrams, and in general there should be a substantial
cancellation among these terms to make the net \nlo{5}
contribution small, as dictated by chiral symmetry. Thus, there
are important differences between $\vA^a_{II}$ of CRSW91 and
$\vA(2\pi)$ derived in $\chi$PT.

\section{Axial Charge Operators}
\setcounter{equation}{0}
\renewcommand{\theequation}{\mbox{B.\arabic{equation}}}
As stressed in the main text, the axial charge operators are
chiral-protected.
Since the axial-charge operators up to ${\cal
O}(Q^4)$ have already been described in detail in
Ref.~\cite{axialch}, we only briefly recapitulate what is directly
relevant to the present work. The leading one-body $A^0$ operator
is kinematically suppressed because of the $\gamma_5$ matrix.
Correspondingly, in $\chi$PT, the $A^0$ operator at order ${\cal
O}(Q^1)$ appears as a $1/m_N$ term. The leading correction to the
one-body axial-charge operator comes from the soft
one-pion-exchange, which is of ${\cal O}(Q^2)$. Loop contributions
start at ${\cal O}(Q^4)$, and hence the ratio of the loop
contribution to the tree diagram two-body contribution is ${\cal
O}(Q^2)$. Finally, there is no three-body contribution up to
${\cal O}(Q^4)$.

The one-body axial-charge operator at threshold is given by \be
A_l^{0,a} &=&
 - \frac{\tau_l^a}{2}\, g_A\,\left[
  \frac{\vs_l\cdot \vbp_l}{m_N}
  + {\cal O}\left(\frac{\vq^2}{m_N^2}\right)\right],
\label{A1B0non} \ee which is ${\cal O}(Q^1)$. We note that
there is no relativistic correction of ${\cal O}(\vq)$
to the one-body axial-charge;
this aspect is in sharp contrast
to the GT operator.

The two-body $A^0$ current appears
at ${\cal O}(Q^2)$ (tree diagram)
and at ${\cal O}(Q^4)$ (loop diagrams):
\be A_{12}^{0,a}= - \frac{g_A}{4 f_\pi^2} \left[
 {\cal T}^{a(I)} {\cal W}^{(I)}
 + {\cal T}^{a(II)} {\cal W}^{(II)} \right]
\ee 
where \be {\cal T}^{a(I)}
  &=& i\, (\vtau_1 \times \vtau_2)^a \, \vk\cdot(\vs_1
    + \vs_2),\\
{\cal T}^{a(II)}
  &=& i \,(\vtau_1 + \vtau_2)^a \, \vk\cdot\vs_1
    \times \vs_2,
\ee with $\vk=\vk_2= -\vk_1$.
The one-pion-exchange contribution including the vertex
renormalization (loop corrections to the vertices) reads \be
{\cal W}^{(I)}_{1\pi}&=& - \frac{1}{m_\pi^2 - t} F_1^V(t),
\nonumber \\
{\cal W}^{(II)}_{1\pi}&=& 0\,, \label{msea} \ee where
$t\equiv k_0^2 - \vk^2 \simeq - \vk^2$,
and $F_1^V(t)$ is the isovector Dirac form
factor of the nucleon electromagnetic current. The
phenomenologically determined $F_1^V(t)$ is of the dipole type \be
F_1^V(t) = \left(\frac{\Lambda^2}{\Lambda^2 - t}\right)^2 \ee with
$\Lambda= 840\ \MeV$. The HB$\chi$PT expression for $F_1^V(t)$ up to
one-loop accuracy is given by
 \be && F_1^V(t) = 1 +
\frac{c_3^R}{f_\pi^2} t \nonumber\\
&&- \frac{t}{16\pi^2f_\pi^2} \left[ \frac{1+3g_A^2}{2} K_0(t)-
2(1+2g_A^2) K_2(t)\right]. \label{F1V} \ee The loop functions,
$K$'s, will be specified below. The constant $c_3^R$ is determined
by the nucleon charge radius,\footnote{ Note that this $c_3^R$ has
nothing to do with the $c$'s that appear in the GT operators. This
confusing situation arises due to the lack of a unified system for
labelling the coefficients in the chiral Lagrangian. In a chiral
expansion ${\cal L}= {\cal L}_0 + {\cal L}_1 + {\cal L}_2 +
\cdots$, it was once common to use the letter $b$ ($c$) to label
the coefficients appearing in ${\cal L}_1$ (${\cal L}_2$). Later,
however, it has become more common to use the letter $c$ to
designate the coefficients that feature in ${\cal L}_1$, resulting
in unfortunate confusions. In short, the above $c_3^R$ is a
coefficient in ${\cal L}_2$, while the $c$'s appearing in the
expression for the GT operator belong to ${\cal L}_1$.} \be c_3^R
\frac{m_\pi^2}{f_\pi^2} = \frac{m_\pi^2}{6} \langle r^2\rangle_1^V
\simeq 0.04784. \label{c3r} \ee We should mention here that ${\cal
M}_{1\pi}$ in Eq.(\ref{msea}) contains both ${\cal O}(Q^2)$ and
${\cal O}(Q^4)$ contributions. The ${\cal O}(Q^2)$ contributions
can be obtained by replacing $F_1^V(t)$ by 1, while $(F_1^V(t)-1)$
is responsible for the ${\cal O}(Q^4)$ contributions. 
The two-pion-exchange contributions,
which are also of ${\cal O}(Q^4)$, are given by :
\be {\cal W}^{(I)}_{2\pi} &=& \frac{1}{16\pi^2 f_\pi^2}
\left[-\frac{3g_A^2-2}{4}K_0(k^2) - \frac12 g_A^2
K_1(k^2)\right]\nonumber\\
 &&- \frac{1}{4 f_\pi^2} \kappa_4^{(1)},
\nonumber \\
{\cal W}^{(II)}_{2\pi} &=& \frac{1}{16\pi^2 f_\pi^2}
 \left[2 g_A^2 K_0(k^2)\right]
 - \frac{1}{4 f_\pi^2} \kappa_4^{(2)}
\label{mab}\ee where $\kappa_4^{(1)}$ and $\kappa_4^{(2)}$ are
unknown parameters\footnote{In fact, these are the only unknown
parameters up to \nlo4. It turns out that due to the chiral filter
mechanism, one can drop these terms when working to \nlo4 for the
GT operator.}. The total two-body axial-charge operator is the sum
of Eqs.~(\ref{msea}) and (\ref{mab}):
\be
{\cal W}^{(I)}={\cal W}^{(I)}_{1\pi}+
{\cal W}^{(I)}_{2\pi}\nonumber\\
{\cal W}^{(II)}={\cal W}^{(II)}_{1\pi}+
{\cal W}^{(II)}_{2\pi}
\ee
The loop functions $K$'s in the
above are defined as \be K_0(t) &=& \int_0^1 dz\, {\rm
ln}\left[1-z(1-z) \frac{t}{m_\pi^2}\right],
\nonumber \\
K_1(t) &=& \int_0^1 dz \frac{-z(1-z) t}{m_\pi^2 - z(1-z)t},
\nonumber \\
K_2(t) &=& \int_0^1 dz \,z(1-z)\, {\rm ln}\left[1-z(1-z)
\frac{t}{m_\pi^2} \right]. \label{Kdef} \ee The integrations over
$z$ can be done analytically, resulting in \be
 K_0(t) &=& -2 + \sigma \,
{\rm ln}\left(\frac{\sigma+1}{\sigma-1}\right),
\nonumber \\
K_1(t) &=& 1-\frac{\sigma^2-1}{2\sigma} \, {\rm
ln}\left(\frac{\sigma+1}{\sigma-1}\right),
\nonumber \\
K_2(t) &=&
 -\frac 49 + \frac{\sigma^2}{6} + \frac{\sigma(3-\sigma^2)}{12}
{\rm ln}\left(\frac{\sigma+1}{\sigma-1}\right), \label{K} \ee with
\be \sigma \equiv \left(\frac{4 m_\pi^2
-t}{-t}\right)^{\frac12}\,. \ee
\section{Regularization and the Cutoff}
\setcounter{equation}{0}
\renewcommand{\theequation}{\mbox{C.\arabic{equation}}}
\subsection{Fourier transform}
\indent Since SNPA employs coordinate-space representation, we
need to Fourier transform the momentum-space expressions. In doing
so, we must impose a cutoff to regularize the integral. The cutoff
introduced here typically represents a scale that divides the
low-energy degrees of freedom (which we choose to include
explicitly) and the high-energy degrees of freedom (which we
integrate out). How to implement cutoff into the theory is not
unique, but physics should be independent of methods used insofar
as the calculation is done consistently. This is a statement of
renormalization group invariance. We now describe a particular
cutoff scheme to be used here.\footnote{ We illustrate our method
for $\vA$, but the same method is used for the other currents as
well.} For the $n_B$-body current in momentum space,
$\vA_{12\cdots n}^{a}= \vA_{12\cdots
n}^{a}(\vk_1,\,\vk_2,\,\cdots,\, \vk_n)$, define its ``Fourier
transform" as \be &&\tilde{\vA}_{12\cdots n}^{a} \equiv
\left[\prod_{l=1}^n \int \!\frac{d^3\vk_l}{(2\pi)^3}\, \e^{i
\vk_l\cdot \vr_l}\, S_\Lambda(\vk_l^2) \right]\,
\nonumber\\
&& (2\pi)^3 \delta^{(3)}(\vq+ \vk_1 + \vk_2 +\cdots + \vk_n)\,
 \vA_{12\cdots n}^a,
\label{FTgene} \ee where $S_\Lambda(\vk^2)$ is a regulator with a
cutoff $\Lambda$, and the factor $(2\pi)^3 \delta^{(3)}(\vq+ \vk_1
+ \vk_2 +\cdots + \vk_n)$ comes from momentum conservation. We
employ here a regulator of the Gaussian type~\footnote{
Using a Gaussian cutoff can in principle 
upset the chiral counting even when graphs 
up to a definite chiral order (say $\nu$)
are considered.  This counting mismatch, however,
occurs at an order higher than $\nu$,
and furthermore the error 
committed is likely to be minimized 
by the regularization scheme
we have adopted for the $\hat{d}^R$ counter term.}
 \be S_\Lambda(\vk^2)
= \exp\left(- \frac{\vk^2}{2 \Lambda^2}\right). \label{regulatorp}
 \ee
For a one-body operator, the regulator plays no role,
see Eq.(\ref{A1Bnon}).
Now for the two-body current,
Eq.(\ref{FTgene}) gives\footnote{ {}From here on, we shall
specialize ourselves to the threshold limit, $\vq=\vzero$.}
 \be
\tilde{\vA}_{12}^{a} =&& \int
\!\frac{d^3\vk}{(2\pi)^3}\, S_\Lambda^2(\vk^2)\, \e^{- i
\vk\cdot \vr_{12}}\nonumber\\
&&\times \vA_{12}^{a}(\vk_1=-\vk,\,\vk_2=\vk)\,.
\label{FTgene2B} \ee
 To simplify subsequent expressions, we define
the following functions: \be \delta_\Lambda^{(3)}\!(\vr) &\equiv&
 \int \!\frac{d^3\vk}{(2\pi)^3}\,
S_\Lambda^2(\vk^2)\, \e^{ i \vk\cdot \vr},
\nonumber \\
y_{0\Lambda}(m,r) &\equiv&
 \int \!\frac{d^3\vk}{(2\pi)^3}\,
S_\Lambda^2(\vk^2)\, \e^{ i \vk\cdot \vr}\, \frac{1}{\vk^2 + m^2},
\nonumber \\
y_{1\Lambda}(m,r) &\equiv& - \frac{1}{m} \frac{\del}{\del r}
y_{0\Lambda}(m,r),
\nonumber \\
y_{2\Lambda}(m,r) &\equiv& \frac{1}{m^2} r \frac{\del}{\del r}
\frac{1}{r} \frac{\del}{\del r} y_{0\Lambda}(m,r).
 \ee
These functions become the ordinary delta and Yukawa functions
when $\Lambda$ goes to infinity. We also use the abbreviation
$y_{0\Lambda}^\pi(r) \equiv y_{0\Lambda}(m_\pi, r)$, and similarly
for $y_{1\Lambda}^\pi(r)$ and $y_{2\Lambda}^\pi(r)$. The
regularized delta and Yukawa functions read
 \be
\delta_\Lambda^{(3)}(\vr) &=& \frac{\Lambda^3}{(4\pi)^{\frac32}}
 \, \exp\left(-\frac{\Lambda^2 r^2}{4}\right),
\label{regdelta}\\
y_{0\Lambda}(m,r) &=&
 \frac{1}{4\pi r} \e^{\frac{m^2}{\Lambda^2}} \frac12
 \left[ \e^{-mr} \mbox{erfc}\left( - \frac{\Lambda r}{2} +
 \frac{m}{\Lambda}\right)\right.\nonumber\\
  &&\left. - (r \leftrightarrow -r) \right]\,.
 \ee

We are now ready to write down the $\nu$=3 two-body axial-vector
current, Eq.(\ref{vAnu3}) in coordinate space:
 \be &&\tilde
\vA_{12}^{a:\nu3}(1\pi) =
\nonumber \\
&& \frac{g_A}{2 m_N f_\pi^2}
\delta_\Lambda^{(3)}(\vr_{12})
\left[
  \frac13 \hat c_3 ( \voO_+^a + \voO_-^a)
 + \frac23 (\hat c_4 + \frac14) \voO_\times^a
\right]
\nonumber \\
&&- \frac{g_A m_\pi^2}{2 m_N f_\pi^2} \Bigg[
 \vokin^a y_{1\Lambda}^\pi(r_{12})
\nonumber \\
&&\ +
 \left( \hat c_3 ( \voT_+^a + \voT_-^a)
 - (\hat c_4 + \frac14) \voT_\times^a
\right)
     y_{2\Lambda}^\pi(r_{12})
\nonumber \\
&&\ + \left(\frac13 \hat c_3
  ( \voO_+^a + \voO_-^a)
 + \frac23 (\hat c_4 + \frac14)
 \voO_\times^a \right) y_{0\Lambda}^\pi(r_{12})
\Bigg],
\\
&&\tilde \vA_{12}^{a:\nu3}(2\pi)=\nonumber\\
 && \frac{g_A}{2 m_N f_\pi^2}\,
 \delta_\Lambda^{(3)}(\vr_{12})
 \left[ \hat d_1 (\voO_+^a + \voO_-^a)
 + 2 \hat d_2 \voO_\times^a
 \right]\,
\label{vAnu3FT} \ee where the superscript $(i,\,j)$ are particle
indices, $r_{12}\equiv |\vr_{12}|$ and $\hatr_{12} \equiv
\frac{\vr_{12}}{r_{12}}$.
In the above equations, we have defined
the following two-body
spin-isospin operators
 \be \oO_{\odot}^{i,a} &\equiv& (\vtau_1 \odot
\vtau_2)^a (\vs_1 \odot \vs_2)^i, \nonumber \\
 \oT_{\odot}^{i,a}
&\equiv& \left(\hat r_{12}^i \hat r_{12}^j - \frac{\delta^{ij}}{3}
\right) \oO_{\odot}^{aj} \ee where $\odot = (+,\,-,\,\times)$ and
\be \okin^{i,a} \equiv - \frac{1}{2 m_\pi} (\vtau_1\times
\vtau_2)^a
   (\bar p_1^i \,\vs_2\cdot\hatr_{12} +
   \bar p_2^i\,\vs_1\cdot\hatr_{12}).
 \ee
Note that $\okin^{i,a}$
is completely determined by Lorentz invariance.
In terms of these seven operators,
we can write {\em all} the two-body currents (including $\nu$=3
and $\nu$=4 contributions) as
 \be \tilde A_{12}^{i,a} &=&
- \sum_{\odot = +,-,\times} \left[ F_\odot^C(r_{12}) \oO_\odot^{i,a}
 + F_\odot^T(r_{12}) \oT_\odot^{i,a} \right]\nonumber\\
 && - \frac{g_A
m_\pi^2}{2 m_N f_\pi^2} y_{1\Lambda}^\pi(r_{12})
\okin^{i,a}\nonumber
\\
&+& \frac{g_A}{2 m_N f_\pi^2}\, \delta_\Lambda^{(3)}(\vr_{12})
 \left[ \sum_{\odot=+,-,\times}
 \hat d_\odot \oO_\odot^{i,a}\right].
\label{vAnuFTp}
 \ee
We have separated out here the part proportional to
$\delta^{(3)}_\Lambda\!(\vr)$. The dimensionless parameters
$\hat d_\odot$ are given by $\hat d_{1,2}$ and (higher order) loop
contributions as \be \hat d_+ &\equiv& \hat d_1 + \frac13 \hat
c_3\nonumber\\
&& - \frac{g_A^2 m_N m_\pi}{32\pi}
  \left(-3 X_1 + 9 X_{\sigma\tau} + 9 X_\sigma - 3 X_\tau\right),
\nonumber \\
\hat d_- &\equiv& \hat d_1 + \frac13 \hat c_3\nonumber\\
 &&- \frac{g_A^2 m_N m_\pi}{32\pi}
  \left(-3 X_1 + 9 X_{\sigma\tau} - 9 X_\sigma + 3 X_\tau\right),
\nonumber \\
\hat d_\times &\equiv&
 2 \left[ \hat d_2 + \frac13 \left(\hat c_4+\frac14\right)
 + \frac{g_A^2 m_N m_\pi}{32\pi}
  X_{\sigma\tau} \right] .
 \ee
The functions $F^{C,T}_\odot$ in Eq.(\ref{vAnuFTp}) are given by
 \be F_+^C(r) &=& \frac{g_A m_\pi^2}{2 m_N f_\pi^2}
  \, \frac{\hat c_3^R}{3} y_{0\Lambda}^\pi(r)
  \nonumber \\
  &+& \frac{g_A^3}{32\pi f_\pi^4} \left\{
   \frac16 (3 D_1 + k^2 D_2-\tilde D_{1\pi})
   \frac{m_\pi^2}{m_\pi^2+k^2}\right.\nonumber\\
   &&\left. + \frac{1}{8} (3 D_1 + k^2 D_2)
   \right. \nonumber \\ && \left.
   +\  \frac{g_A^2}{64}\left(3D_3 - k^2 D_4 + 2 k^2 D_5\right)
   \right\}_{\rm FT}(r),
\nonumber \\
F_-^C(r) &=& \frac{g_A m_\pi^2}{2 m_N f_\pi^2}
  \, \frac{\hat c_3^R}{3} y_{0\Lambda}^\pi(r)
  \nonumber \\
  &+& \frac{g_A^3}{32\pi f_\pi^4} \left\{
   \frac16 (3 D_1 + k^2 D_2-\tilde D_{1\pi})
    \frac{m_\pi^2}{m_\pi^2+k^2}\right.\nonumber\\
   &&\left. + \frac{1}{24} (3 D_1 + k^2 D_2)
   \right. \nonumber \\ && \left.
   +\ \frac{g_A^2}{64}\left(-D_3 + \frac13 k^2 D_4 +
   \frac{2}{3} k^2 D_5\right)
   \right\}_{\rm FT}(r),
\nonumber \\
F_\times^C(r) &=& \frac{g_A m_\pi^2}{2 m_N f_\pi^2}
  \, \frac23 \left(\hat c_4^R + \frac14\right)
  y_{0\Lambda}^\pi(r)
+ \frac{g_A^3}{32\pi f_\pi^4} \nonumber\\
&&\otimes\left\{
   \frac23 D_1
   -\frac23 (D_1-D_{1\pi})\frac{m_\pi^2}{m_\pi^2+k^2}
   \right\}_{\rm FT}(r)
\ee and
 \be F_+^T(r) &=& \frac{g_A m_\pi^2}{2 m_N f_\pi^2}
  \, \hat c_3^R y_{2\Lambda}^\pi(r)
  \nonumber \\
  &+& \frac{g_A^3}{32\pi f_\pi^4} \left\{
   \frac12 (3D_1 + k^2 D_2-\tilde D_{1\pi})
   \frac{1}{m_\pi^2+k^2}
   \right.\nonumber\\
   &&\left. + \frac14 D_2
   + \frac{g_A^2}{64}\left(-6 D_4 + 3 D_5\right)
   \right\}_{\rm FT}(r),
\nonumber \\
F_-^T(r) &=& \frac{g_A m_\pi^2}{2 m_N f_\pi^2}
  \, \hat c_3^R y_{2\Lambda}^\pi(r)
  \nonumber \\
  &+& \frac{g_A^3}{32\pi f_\pi^4} \left\{
   \frac12 (3 D_1 + k^2 D_2-\tilde D_{1\pi})
    \frac{1}{m_\pi^2+k^2}\right.\nonumber\\
&&\left.   - \frac14 D_2
   + \frac{g_A^2}{64}\left(2 D_4 + D_5\right)
   \right\}_{\rm FT}(r),
\nonumber \\
F_\times^T(r) &=& - \frac{g_A m_\pi^2}{2 m_N f_\pi^2}
  \, \left(\hat c_4^R + \frac14\right) y_{2\Lambda}^\pi(r)
 + \frac{g_A^3}{32\pi f_\pi^4} \nonumber\\
 && \times \left\{
   \frac{D_1-D_{1\pi}}{m_\pi^2+k^2}
   \right\}_{\rm FT}^T(r),
\ee where $k\equiv |\vk|$, $D_i=D_i(k)$ and \be \left\{ f(k^2)
\right\}_{\rm FT}(r) &\equiv& \int \!\frac{d^3\vk_2}{(2\pi)^3}\,
S_\Lambda^2(\vk^2)\, \e^{- i \vk\cdot \vr}\, f(k^2),
\nonumber \\
\left\{ f(k^2) \right\}_{\rm FT}^T(r) &\equiv& r \frac{\del}{\del
r} \frac{1}{r} \frac{\del}{\del r} \int
\!\frac{d^3\vk_2}{(2\pi)^3}\, S_\Lambda^2(\vk^2)\, \e^{- i
\vk\cdot \vr}\, f(k^2).\nonumber
 \ee
The explicit results of Fourier transformation of $D_i(k)$ and
$k^2 D_i(k)$ are given by (``$\rightarrow$" denotes Fourier
transformation):
 \be D_1 &\rightarrow&
 -\frac{m_\pi^2}{2\pi r^2} \zint K_2(x),
\nonumber \\
D_2 &\rightarrow&
 \frac{1}{2\pi r^2} \zint \zz\, x K_1(x),
\nonumber \\
D_3 &\rightarrow&
 \frac{m_\pi^2}{2\pi r^2} \zint (6 K_2(x) + x K_1(x)),
\nonumber \\
D_4 &\rightarrow&
 \frac{1}{2\pi r^2} \zint \left[
 \zz\,x^2 K_0(x) + (6 \zz -1) x K_1(x)\right],
\nonumber \\
D_5 &\rightarrow&
 \frac{1}{2\pi r^2} \zint x K_1(x)
\label{Dix}\ee and \be \vk^2 D_2 &=& D_1 - m_\pi^2 D_5 \rightarrow
 - \frac{m_\pi^2}{2\pi r^2} \zint (K_2(x) + x K_1(x)),
\nonumber \\
\vk^2 D_4 &\rightarrow&
 -\frac{m_\pi^2}{2\pi r^2} \zint \left[
 \frac{6\zz -1}{\zz} (2 x K_1(x)+ K_2(x))\right.\nonumber\\
 &&\left. + x^2 K_0(x) - x K_1(x)\right],
\nonumber \\
\vk^2 D_5 &\rightarrow&
 -\frac{m_\pi^2}{2\pi r^2} \zint
  \frac{1}{\zz} (2 x K_1(x)+ K_2(x)),
\label{vq2Di}\ee where \be x \equiv \frac{m_\pi r}{\sqrt{\zz}}.
\ee

Next we turn to the three-body currents of Eq.~(\ref{AIII}). The
Fourier-transformed three-body current has the form
 \be &&\tilde
\vA_{123}^{a}\nonumber\\ &&= -\sum_{{\rm cycle}(123)}\frac{g_A^3}{16
f_\pi^4}
 (2 \tau_1^a\, \vtau_2\cdot\vtau_3
 -  \tau_2^a\, \vtau_3\cdot\vtau_1
 - \tau_3^a\, \vtau_1\cdot\vtau_2)\,
 \tilde \vI_{123}\nonumber\\
\ee with
 \be \tilde \vI_{123} &\equiv& \left[\prod_{i=l}^3 \int
\!\frac{d^3\vk_l}{(2\pi)^3}\, \e^{i \vk_l\cdot \vr_l}\, \e^{-
\frac{\vk_l^2}{2 \Lambda^2}}\right]\nonumber\\
&&\otimes (2\pi)^3 \delta^{(3)}(\vk_1 + \vk_2 + \vk_3)\,
\nonumber \\
&&\
 \left( \vs_1 - \frac{4}{3} \frac{\vk_1\, \vs_1\cdot \vk_1}{\vk_1^2+m_\pi^2}
 \right)
 \frac{\vs_2\cdot \vk_2}{\vk_2^2+m_\pi^2}
 \frac{\vs_3\cdot \vk_3}{\vk_3^2+m_\pi^2}\,.\nonumber
 \ee The calculation of $\tilde I^i_{123}$ is rather
involved. We may start with exploiting the identity
 \be (2\pi)^3
\delta^{(3)}(\vk_1+\vk_2+\vk_3)
 = \int d^3\vx\, \e^{i \vx \cdot (\vk_1+\vk_2+\vk_3)}
\ee to arrive at
\be \tilde I^i_{123} &=& - \frac{4 m_\pi^4}{3} \int \!d^3\vx\,
\left[ \frac{5}{12 m_\pi^2} \sigma_1^i \delta^{(3)}_\bLambda(\vx)
+ \sigma_1^i y_{0\bLambda}^\pi(|\vx|) \right.\nonumber\\
&&\left. - \left(\hat x^i \hat x^j - \frac{\delta^{ij}}{3}\right)
      \sigma_1^j y_{2\bLambda}^\pi(|\vx|) \right]
\frac{\vs_2 \cdot (\vx+\vr_{12})}{|\vx+\vr_{12}|}\,
\nonumber \\
&\otimes& y_{1\bLambda}^\pi(|\vx+\vr_{12}|)\, \frac{\vs_3 \cdot
(\vx+\vr_{13})}{|\vx+\vr_{13}|}\,
y_{1\bLambda}^\pi(|\vx+\vr_{13}|) ,\nonumber
 \ee where
 \be \bLambda \equiv
\sqrt{2} \Lambda\,.\nonumber \ee
This representation is nicely transparent, and the resulting
integrand is non-oscillatory and rapidly damping. The remaining
integration can be done by means of a Monte Carlo simulation with
Metropolis random walks.

\subsection{The parameter $\hat d^R$}
\indent Up to \nlo3, unknown parameters occur only in $\vA$. At
\nlo4, several unknown parameters appear in both $\vV$ and $A_0$,
but no new parameters appear in $\vA$. By the chiral filter
argument, one can ignore the \nlo4 terms in both $\vV$ and $A_0$
while going to \nlo4 in $\vA$. Thus, up to \nlo4, the only
genuinely unknown parameters reside in $\vA$. The crucial
observation is that {\it up to \nlo4, there is effectively only
one constant $\hat d^R$ that governs the GT amplitudes of all the
cases under consideration.} The argument goes as follows.

The two parameters, $\hat d_1$ and $\hat d_2$, and the four $X$'s
can be combined into three unknown parameters, $\hat
d_{\pm,\times}$, that reflect short-range physics. It is the
Fermi-Dirac statistics that reduces the number of unknowns from
three to one.  To see this, let $\Xi^\sigma$ and $\Xi^\tau$ be the
exchange operators in spin and isospin spaces, respectively;
$\Xi^\sigma= \frac12 (1 + \vs_1 \cdot \vs_2)$,
and $\Xi^\tau= \frac12 (1 + \vtau_1 \cdot \vtau_2)$.
An explicit calculation gives the identity
$\vs_1\times \vs_2 = i (\vs_1- \vs_2) \Xi^\sigma$, and likewise
for $\vtau_1\times \vtau_2$. Now, the Fermi-Dirac statistics
requires that $\Xi^r \Xi^\sigma \Xi^\tau= -1$, where $\Xi^r$ is
the Majorana exchange operator that exchanges the orbital
coordinates, $\vr_1$ and $\vr_2$. As a result, \be
\oO_\times^{i,a} = - \oO_-^{i,a} \Xi^\sigma \Xi^\tau = \oO_-^{i,a}
\Xi^r . \ee When multiplied by the delta function
$\delta^{(3)}(\vr)$, the operators are non-vanishing only for the
$L=0$ states, which then implies $S+T=1$. Acting on $L=0$ states,
$\oO_+^{i,a}$ is identically zero, since either spin or isospin
must be equal to zero. Furthermore, the $L=0$ states are
eigenstates of the operator $\Xi^r$ with eigenvalue 1, so that
$\oO_-^{i,a}$ becomes identical to $\oO_\times^{i,a}$. Thus we are
left with only one unknown parameter, $\hat d^R \equiv \hat d_- +
\hat d_\times$.

The above argument is not strictly valid for the cut-off delta
function $\delta_\Lambda^{(3)}(\vr)$, which has a finite (albeit
very small) range, $\sim \Lambda^{-1}$. However, deviations from
the ordinary delta function case is higher order in chiral
counting and hence can be ignored.

\section{Calculation up to \nlo3}\label{n3lo}
\setcounter{equation}{0}
\renewcommand{\theequation}{\mbox{D.\arabic{equation}}}
We have derived all the weak currents up to \nlo4. As Table
\ref{power} indicates, loop contributions start at \nlo4. Loop
corrections in the vector currents (both $\vV$ and $V^0$) can
be safely ignored, since even their leading single-particle terms
are suppressed relative to the axial current. It turns out that
the loop diagrams in $\vA$ are all finite and hence need no
regularization although there are finite counter terms that should
be taken into account. On the other hand, the loop diagrams in
$A_0$ do have divergences and need to be regularized. To derive
the momentum space expressions for the currents given above, we
have employed the dimensional regularization. This is not quite
congruous with the cutoff regularization adopted in going from
momentum to coordinate space. Meanwhile, using a cutoff
regularization in evaluating loop graphs is a delicate matter,
since that might endanger chiral symmetry; with the use of a
cutoff regularization one might need chiral-symmetry-breaking
counter terms in order to satisfy the Ward identities. We have not
yet investigated whether the dimensional regularization as used
here preserves chiral symmetry, and hence we cannot say at this
point whether our coordinate space operators at \nlo4 are
fully consistent.
However, this problem does not arise
if we limit ourselves to \nlo3, for up
to this order there are no loop contributions. The relevant 2-body
currents in coordinate space are: \be \vV_{12}(\vr) &=& -
\tau_\times^- \frac{g_A^2 m_\pi^2}{12 f_\pi^2} \vr \,
 \left[
 \vs_1\cdot\vs_2 y_{0\Lambda}^\pi(r)
 + S_{12}(\hat{r}) y_{2\Lambda}^\pi(r) \right]
\nonumber \\
 &-& i \frac{g_A^2}{8 f_\pi^2} \vk\times \left[
 \voO_\times
 y_{0\Lambda}^\pi(r)
 +\ \left( \voT_\times -
 \frac23 \voO_\times \right)
  y_{1\Lambda}^\pi(r)
 \right],
\nonumber \\
A^{0}_{12}(\vr) &=& \tau_\times^- \frac{g_A}{4 f_\pi^2}
\left[
\vs_+ \cdot \hat{r} \frac{y_{1\Lambda}^\pi(r)}{r} + \frac{i}{2}
\vk\cdot \hat{r}\, \vs_-\cdot \hat{r}\,y_{1\Lambda}^\pi(r)
\right],
\nonumber \\
\vA_{12}(\vr) &=& - \frac{g_A m_\pi^2}{2 m_N f_\pi^2} \left[
\frac{}{}\right.
\nonumber \\
&&\left.
  \left\{
   \frac{\hat c_3}{3} (\voO_+ + \voO_-)
  + \frac23 \left(\hat c_4 + \frac14\right)
   \voO_\times \right\} y_{0\Lambda}^\pi(r)
\right. \nonumber \\
&&\left.
  + \left\{
     \hat c_3 (\voT_+ + \voT_-)
  - \left(\hat c_4 + \frac14\right) \voT_\times
     \right\} y_{2\Lambda}^\pi(r)
  \right]
\nonumber \\
&+& \frac{g_A}{2 m_N f_\pi^2 r}
\left[
\frac{1}{2} \tau_\times^-
   (\vbp_1 \,\vs_2\cdot\hatr +
   \vbp_2\,\vs_1\cdot\hatr)
y_{1\Lambda}^\pi(r)
\right.
\nonumber \\
&&\left.
 + \delta_\Lambda^{(3)}(\vr)
 \, \hat d^R \voO_\times \right],
\label{2bodynlo3}\ee
 with \be \left(\delta_\Lambda^{(3)}(\vr),\
y_{0\Lambda}^\pi(r)\right) \equiv
 \int\!\!\frac{d^3\vk}{(2\pi)^3}\,
 \e^{-\frac{\vk^2}{\Lambda^2}}
\e^{i \vk\cdot \vr} \left(1,\ \frac{1}{\vk^2 + m_\pi^2}\right),
\nonumber \ee
$y_{1\Lambda}^\pi(r) \equiv - r \frac{\del}{\del r}
y_{0\Lambda}^\pi(r)$ and $y_{2\Lambda}^\pi(r) \equiv
\frac{r}{m_\pi^2} \frac{\del}{\del r} \frac{1}{r} \frac{\del}{\del
r} y_{0\Lambda}^\pi(r)$,
where $ \oO_{\odot}^{k} \equiv \tau_\odot^- \vs_\odot^k$, $
\oT_{\odot}^k \equiv \left(\hat r^k \hat r^\ell -
\frac{\delta^{k\ell}}{3} \right) \oO_{\odot}^{\ell}$ and \be
\dR\equiv \hat d_1 +2 \hat d_2 + \frac13 \hat c_3
 + \frac23 \hat c_4 + \frac16\,.
 \label{parameter}
\ee The derivative operators, $\vbp_l$ ($l=1,\,2$) in
eq.(\ref{vAnuFT}), should be understood to act only on the wave
functions.

Since ${\hat d}_R$ accompanies the regularized $\delta$-function
$\delta_\Lambda^{(3)}(\vr)$, its contribution depends on $\Lambda$
rather strongly. However, the renormalization-group invariance of
EFT requires that this sensitivity to $\Lambda$ should be
compensated by the contributions of the remaining terms. Since the
single-particle piece of $\vA$ has no $\Lambda$ dependence,
and since all the currents other than $\vA$ have only weak
$\Lambda$ dependence, this compensation must occur between the
finite-range two-body GT and the regularized delta-function term.
This has been indeed verified in our calculation over a wide range
of $\Lambda$ (500 MeV -  800 MeV) although,
as mentioned above, the results for the 800
MeV cutoff should be viewed with caution.

\bi{doubledecimation} G.E. Brown and M. Rho, ``Double decimation
and sliding vacua in the nuclear many-body system," to appear.

\bi{harada-yamawaki} M. Harada and K. Yamawaki, \pr \ {\bf D64},
014023 (2001); Phys. Rep., to appear.

\bi{BR} G.E. Brown and M. Rho, \prl \ {\bf 66}, 2720 (1991).

\bi{kuo} S.K. Bogner, T.T.S. Kuo, A. Schwenk, D.R. Entem and R.
Machleidt, ``Towards a unique low momentum nucleon-nucleon
interaction", nucl-th/0108041; S.K. Bogner, A. Schwenk, T.T.S. Kuo,
and G.E. Brown, ``Renormalization group equation for low momentum
effective nuclear interactions", nucl-th/0111042.

\bi{schwenk} A. Schwenk, B. Friman, and G.E. Brown, \np \ {\bf
A703}, 745 (2002).

\bibitem{PMetal2001}
T.-S. Park, L.E. Marcucci, R. Schiavilla, M. Viviani, A. Kievsky,
S. Rosati, K. Kubodera, D.-P. Min, and M. Rho, nucl-th/0106025.

\bibitem{PMetal2}
T.-S. Park, L.E. Marcucci, R. Schiavilla, M. Viviani, A. Kievsky,
S. Rosati, K. Kubodera, D.-P. Min, and M. Rho, nucl-th/0107012.

\bi{controversy}
J.N. Bahcall, Phys. Rep. {\bf 333}, 47 (2000),
 and references therein.

\bi{monderen}
R. Escribano, J.-M. Frere, A. Gevaert, and D.
Monderen, Phys. Lett. {\bf B444}, 397 (1998).

\bi{KDR}
K. Kubodera, J. Delorme, and M. Rho, \prl \ {\bf 40}, 755
(1978);  M. Rho, \prl\ {\bf 66}, 1275 (1991).

\bi{BR2001} G.E. Brown and M. Rho,  Phys. Rep. {\bf 363}, 85
(2002); hep-ph/0103102.

\bi{ananyan} S.M. Ananyan, B.D. Serot and J.D. Walecka, ``The
axial-vector current in nuclear many-body physics,"
nucl-th/0207019.

\bibitem{M1}
T.-S. Park, D.-P. Min, and M. Rho, Phys.
Rev. Lett. {\bf 74}, 4153 (1995);
Nucl. Phys. {\bf A596}, 515 (1996).

\bibitem{axialch}
T.-S. Park, D.-P. Min, and M. Rho,
Phys. Rep. {\bf 233}, 341 (1993);
T.-S. Park, I.S. Towner, and K. Kubodera,
Nucl. Phys. {\bf A579}, 381 (1994).

\bibitem{CRSW91}
J. Carlson, D.O. Riska, R. Schiavilla, and R.B. Wiringa, Phys.
Rev. {\bf C44}, 619 (1991).

\bibitem{SWPC92}
R. Schiavilla, R.B. Wiringa, V.R. Pandharipande, and J. Carlson,
Phys. Rev. {\bf C45}, 2628 (1992).

\bibitem{MSVKRB} L.E. Marcucci, R. Schiavilla, M. Viviani, A. Kievsky,
S. Rosati, and J.E. Beacom, \pr \ {\bf C63}, 015801 (2001).

\bi{challenge}
J.N. Bahcall, hep-ex/0002018; J.N. Bahcall and P.I.
Krastev, Phys. Lett. {\bf B436}, 243 (1998).

\bibitem{PKMR}
T.-S. Park, K. Kubodera, D.-P. Min, and M. Rho,
Nucl. Phys. {\bf A684}, 101 (2001); see also nucl-th/9904053.

\bibitem{weinberg}
S. Weinberg, Phys. Lett. {\bf B251}, 288 (1990);
Nucl. Phys. {\bf B363}, 3 (1991);
Phys. Lett. {\bf B295}, 114 (1992).

\bibitem{npp} T.-S. Park, K. Kubodera,
D.-P. Min, and M. Rho, Phys. Lett. {\bf B472}, 232 (2000).

\bibitem{beane}
S.R. Beane, V. Bernard, T.-S.H. Lee, and U.-G. Mei{\ss}ner,
Phys. Rev. {\bf C57}, 424 (1998);
S.R. Beane, V. Bernard, T.-S.H. Lee,
U.-G. Mei{\ss}ner, and U. van Kolck,
Nucl. Phys. {\bf A618}, 381 (1997).

\bibitem{pp}
T.-S. Park, K. Kubodera, D.-P. Min, and M. Rho,
Astrophys. J. {\bf 507}, 443 (1998).

\bibitem{TBDexp}
R. Schiavilla {\it et al.},
Phys. Rev. {\bf C58}, 1263 (1998).

\bibitem{bernard}
V. Bernard, N. Kaiser and U.-G. Mei{\ss}ner, Nucl. Phys.  {\bf
A615}, 483 (1997).

\bibitem{Kolck3}  T. D. Cohen, J. L. Friar, G. A. Miller,
and U. van Kolck, Phys. Rev. {\bf C53}, 2661 (1996).

\bibitem{PDG}
D.E. Groom {\it et al.} (Particle Data Group), Eur. Phys. Jour.
{\bf C15}, 1 (2000).

\bibitem{csTREE}
V. Bernard, N. Kaiser, and U.-G. Mei{\ss}ner, Nucl. Phys. {\bf
B457}, 147 (1995).

\bibitem{roccoetal}
L.E. Marcucci, R. Schiavilla, M. Viviani, A. Kievsky, and S. Rosati,
Phys. Rev. Lett. {\bf 84}, 5959 (2000).

\bibitem{VKR95}
M. Viviani, A. Kievsky, and S. Rosati,
Few-Body Syst. {\bf 18}, 25 (1995).

\bibitem{VRK98}
M. Viviani, S. Rosati, and A. Kievsky,
Phys. Rev. Lett. {\bf 81}, 1580 (1998).

\bibitem{av18}
R.B. Wiringa, V.G.J. Stoks, and R. Schiavilla,
Phys. Rev. {\bf C51}, 38 (1995).

\bibitem{uix}
B.S. Pudliner, V.R. Pandharipande,
J. Carlson, and R.B. Wiringa,
Phys. Rev. Lett. {\bf 74}, 4396 (1995).

\bibitem{KB94}
M. Kamionkowski and J.N. Bahcall,
Astrophys. J. {\bf 420}, 884 (1994).

\bibitem{hardy}
J.C. Hardy, I.S. Towner, V.T. Koslowsky,
 E. Hagberg, and H. Schmeing,
 Nucl. Phys. {\bf A509}, 429 (1990).

\bi{SK2001} S. Fukuda {\it et al.}, Phys. Rev. Lett. {\bf 86},
5651 (2001).

\bi{BP2000} J.N. Bahcall, M.H. Pinsonneault, and S. Basu,
Astrophys. J. {\bf 555}, 990 (2001); astro-ph/0010346.

\bi{PKMRhep}
T.-S. Park, K. Kubodera, D.-P. Min, and M. Rho,
unpublished.

\bibitem{forest96}
J.L. Forest, V.R. Pandharipande, S.C. Pieper, R.B. Wiringa, R.
Schiavilla, and A. Arriaga, Phys. Rev. {\bf C54}, 646 (1996).

\bi{crs99}
J.-W. Chen, G. Rupak, and M.J. Savage, \pl \
{\bf B464}, 1 (1999).

\bi{PKMR98} T.-S. Park, K. Kubodera, D.-P. Min, and M. Rho, \pr \
{\bf C58}, 637 (1998).

\bi{beaneetal2} S.R. Beane, P.F. Bedaque, M.J. Savage, and U. van
Kolck, \np \ {\bf A700}, 377 (2002); nucl-th/0104030.

\bi{kuo2} S. Bogner, T.T.S. Kuo, L. Coraggio, A. Covello and N.
Itaco, ``A new approach to shell model effective interactions
based on low-momentum nucleon-nucleon potential", nucl-th/0108040.

\bibitem{pds}
D.B. Kaplan, M.J. Savage, and M.B. Wise, Phys. Lett. {\bf B424},
390 (1998); Nucl. Phys. {\bf B534}, 329 (1998).

\bi{seattle} S.R. Beane, P.F. Bedaque, W.C. Haxton, D.R. Phillips
and M.J. Savage, in {\it ``At the Frontier of Particle Physics --
 Handbook of QCD"}, ed. by M. Shifman
 (World Scientific, Singapore, 2001),
 Vol.\ 1, p.133

\bibitem{bc01}
X. Kong and F. Ravendal, Nucl. Phys. {\bf A656}, 421 (1999); Nucl.
Phys. {\bf A665}, 137 (2000); Phys. Lett. {\bf B470}, 1 (1999);
nucl-th/0004038; M. Butler and J.-W. Chen, nucl-th/0101017.

\bi{bedaque02}  P. F. Bedaque, H.W. Griesshammer, H.-W. Hammer, and
G. Rupak, ``Low energy expansion in the three body system to all
orders and the triton channel", nucl-th/0207034

\bibitem{epeetal00}
E. Epelbaum, W. Gl\"{o}ckle, and U.-G. Mei{\ss}ner, Nucl. Phys.
{\bf A671}, 295 (2000); Phys. Rev. Lett. {\bf 86}, 4787 (2000); E.
Epelbaum {\it et al.}, nucl-th/0109065, and references therein.

\bibitem{PJM} T.-S. Park, H. Jung, and D.-P. Min,
Phys. Lett. {\bf B409}, 26 (1997).

\bibitem{fettes}  N. Fettes, U.-G. Mei{\ss}ner, and S. Steinberg,
Nucl. Phys. {\bf A 640}, 199 (1998).



\bibitem{cr71}
M. Chemtob and M. Rho, Nucl. Phys. {\bf A163}, 1 (1971).

\bibitem{towner} I.S. Towner, Phys. Rep. {\bf 155}, 263 (1987).

\bi{IT79} E. Ivanov and E. Truhlik, Nucl. Phys. {\bf A316}, 451
(1979); {\it ibid} 437.

\bi{cohen} D.R. Phillips and T.D. Cohen, Nucl. Phys. {\bf A668}, 45
(2000).

\end{document}